\documentclass[10pt, journal, a4paper]{IEEEtran}

\ifCLASSINFOpdf

\else

\fi

\usepackage{mdframed}
\usepackage{xcolor}
\usepackage{CJK}
\usepackage{algorithm}
\usepackage{pifont}
\usepackage{mathrsfs}
\usepackage{algorithmic}
\usepackage{amsmath}
\usepackage{amssymb}
\usepackage{psfrag}
\usepackage{graphicx}
\usepackage{epstopdf}
\usepackage{multirow}
\usepackage{stfloats}
\usepackage{cite}
\usepackage{subcaption}
\usepackage{color}
\usepackage[normalem]{ulem}
\usepackage{arydshln}
\usepackage{enumerate}
\usepackage{bbm}
\usepackage{verbatim}
\usepackage{url}
\usepackage[hidelinks]{hyperref} 
\usepackage[T1]{fontenc}
\usepackage[utf8]{inputenc}

\newtheorem{lemma}{\bf{Lemma}}[section]

\newtheorem{remark}{\bf{Remark}}
\newcommand{\bm}[1]{\mbox{\boldmath{$#1$}}}

\def\BibTeX{{\rm B\kern-.05em{\sc i\kern-.025em b}\kern-.08em
    T\kern-.1667em\lower.7ex\hbox{E}\kern-.125emX}}
\usepackage{balance}

\newcommand{\tabincell}[2]{\begin{tabular}{@{}#1@{}}#2\end{tabular}}

\newmdenv[
  backgroundcolor=white,
  linewidth=0.75pt,
  linecolor=black,
  leftline=true,
  topline=true,
  rightline=true,
  bottomline=true
]{mybox}
\begin{document}
\title{OTFS-MDMA: An Elastic Multi-Domain Resource Utilization  Mechanism for High Mobility Scenarios}

\author{Jie Chen, \IEEEmembership{Member, IEEE}, Xianbin Wang, \IEEEmembership{Fellow, IEEE}, and Lajos Hanzo, \IEEEmembership{Life Fellow, IEEE}
\thanks{ Manuscript submitted to IEEE Journal on Selected Areas in Communications 27 February 2024; revised 30 August 2024;  accepted 6 November 2024. (Corresponding author: Xianbin Wang.)

The work of Jie Chen and Xianbin Wang was supported in part by the Natural Sciences and Engineering Research Council of Canada (NSERC) Discovery Program under Grant RGPIN-2024-05720  and in part by the Canada Research Chair Program.

L. Hanzo would like to acknowledge the financial support of the Engineering and Physical Sciences Research Council (EPSRC) projects under grant EP/Y037243/1, EP/W016605/1, EP/X01228X/1, EP/Y026721/1, EP/W032635/1, and EP/X04047X/1.

 J. Chen and X. Wang are with the  Department of Electrical and Computer Engineering, Western University, London, ON N6A 5B9, Canada (e-mails: jche2426@uwo.ca, xianbin.wang@uwo.ca). L. Hanzo is with the School of Electronics and Computer Science, University of Southampton, Southampton SO17 1BJ, U.K. (e-mail: lh@ecs.soton.ac.uk).
}

}
 \maketitle
\begin{abstract}
By harnessing the delay-Doppler (DD) resource domain, orthogonal time-frequency space (OTFS) substantially
improves the communication performance under high-mobility scenarios by maintaining quasi-time-invariant channel characteristics.
However, conventional multiple access (MA) techniques fail to efficiently support OTFS in the face of diverse communication requirements.
Recently, multi-dimensional MA (MDMA) has emerged as a flexible channel access technique by elastically exploiting multi-domain resources for tailored service provision.
Therefore, we conceive an elastic multi-domain resource utilization mechanism for a novel multi-user OTFS-MDMA system by leveraging user-specific channel characteristics across the DD, power, and spatial resource domains.
Specifically, we divide all DD resource bins into separate subregions called DD resource slots (RSs), each of which supports a fraction of users, thus reducing the multi-user interference.
Then, the most suitable MA, including orthogonal, non-orthogonal, or spatial division MA (OMA/ NOMA/ SDMA), will be selected
with each RS based on the interference levels in the power and spatial domains, thus enhancing the spectrum efficiency.
Then, we jointly optimize the user assignment, MA scheme selection, and power allocation in all DD RSs to maximize the weighted sum-rate subject to their minimum rate and various practical constraints.
Since this results in a non-convex problem, we develop a dynamic programming and monotonic optimization (DPMO) method to find the globally optimal solution in the special case of disregarding rate constraints. Subsequently, we apply a low-complexity algorithm to find sub-optimal solutions in general cases.
\end{abstract}

\section{Introduction}
Next-generation cellular systems are expected to provide reliable communications for a massive number of users~\cite{tataria20216g,zhang20196g,wang2023realizing}.
This is particularly challenging for the dynamic channels encountered in high-mobility scenarios of aircraft-to-ground communications, industrial automation, intelligent transportation, etc.
Conventional multi-carrier modulation techniques, such as orthogonal frequency division multiplexing (OFDM), suffer from performance degradation in high-mobility scenarios due to the doubly selective channels with substantial delay-Doppler spreads \cite{xu2023optical}.
On the other hand, conventional multiple access (MA) schemes such as frequency division, time division, spatial division, or non-orthogonal MA (FDMA, TDMA, SDMA, NOMA) have inherent limitations in terms of accommodating numerous users with high spectrum efficiency. This is because their rigid resource utilization and dependence on a single resource domain result in a lack of flexibility in supporting diverse users.
This motivates us to investigate novel elastic multi-dimensional modulation and access mechanisms to fulfill diverse user demands in mobile communications.

Orthogonal time-frequency space (OTFS) modulation has demonstrated superior performance compared to traditional multi-carrier modulation methods in high-mobility scenarios \cite{mohammed2022otfs,hadani2017orthogonal,reddy2023performance,wei2021orthogonal,sui2023space}.
This is because OTFS maps the data symbols to the two-dimensional delay-Doppler (DD) resource domain, and each path of a doubly-selective channel with specific delay and Doppler spreads can be represented by a specific quasi-time-invariant channel tap on the DD plane.
Consequently, the received OTFS signals can be viewed as a two-dimensional convolution between the information symbols and the effective DD domain channel. This scheme harnesses both the frequency- and time-diversity of doubly-selective channels, enabling reliable communication \cite{lin2022orthogonal}.
Moreover, despite the complexity challenges associated with OTFS \cite{gaudio2021otfs}, its time-invariant channel characteristics in the DD domain facilitate stable resource allocation in mobility scenarios, making it more promising than OFDM facing time-varying channels, particularly in multi-user cases.
This advantage arises because adapting resource allocation for time-varying channels increases both design complexity and control overhead for exchanging updated allocation schemes between transceivers. References \cite{mohammadi2023cell,li2023delay} demonstrate that as the degrees of freedom in the optimization variables increase, the performance gap between OTFS and OFDM becomes more pronounced.

Due to its promising potential, OTFS modulation has garnered significant attention recently from both industry and academia.
Specifically, a simple matrix representation of the input-output relationship of OTFS modulation using practical pulse-shaping waveforms was developed in \cite{raviteja2018practical}.
The resultant relationship exhibits a sparse structure, which facilitates the implementation of sophisticated detection algorithms having low computational complexity.
Then, a simplified linear minimum mean square error (MMSE) receiver was introduced in \cite{tiwari2019low}, which utilizes the sparse nature of the channel and the quasi-band structure of the matrices in the demodulation process to reduce the decoding complexity.
The authors of \cite{raviteja2018interference} introduced an iterative interference canceller and detector that employs message passing for joint inter-Doppler and inter-carrier interference cancellation as well as symbol detection.
Then, this detector was extended in  \cite{yuan2021iterative} by using unitary approximate techniques and exploiting the structured nature of the channel matrix.
Furthermore, the channel sparsity was exploited to devise a Bayesian framework for estimating DD domain channels for single-input single-output (SISO)  \cite{wei2022off} and multiple-input multiple-output (MIMO) \cite{mehrotra2023online} systems.

As expected, there is existing literature investigating OTFS systems relying on various MA technologies.
Specifically,
for single-antenna OTFS systems \cite{khammammetti2022spectral,habibi2024user,ding2019otfs,ge2021otfs,ge2023otfs,ma2021otfs}, the classical orthogonal MA (OMA) methods were investigated in \cite{khammammetti2022spectral,habibi2024user}, where users are assigned distinct delay, Doppler, or frequency resource bins to reduce co-channel interference.
Besides, \cite{khammammetti2022spectral} found that interleaved DD domain OMA offers better spectral efficiency than interleaved time-frequency (TF) domain OMA.
Then,  power-domain NOMA was investigated in \cite{ding2019otfs,ge2021otfs} to enable non-orthogonal spectrum sharing between high-mobility and low-mobility users.
The sophisticated equalization and message-passing algorithms were used in \cite{ding2019otfs} and \cite{ge2021otfs}, respectively, to achieve efficient OTFS signal reception.
In \cite{ge2023otfs}, a sparse code MA (SCMA) scheme was conceived for the uplink of coordinated multi-point vehicular networks, where the performance was improved by leveraging additional diversity from both the Doppler and spatial domains.
Next, a tandem spreading MA (TSMA) arrangement was introduced in \cite{ma2021otfs} for a large-scale network, where segment coding and tandem spreading methods were deployed to improve both connectivity and reliability by sacrificing the data rate.
Later, for multi-antenna OTFS systems \cite{li2020new,liu2023predictive},  path division MA (PDMA) was proposed in \cite{li2020new} for both uplink and downlink by allocating angle-domain resources for multiple users to avoid co-channel interference.
As a further advance, a deep learning-based predictive precoder design was proposed in \cite{liu2023predictive} to enable ultra-reliable low-latency communications.

\begin{table*}[t]{
\renewcommand{\arraystretch}{0.9}
\caption{Recent advances in OTFS and MDMA systems}\label{table00}\vspace{-0.2cm}
 \centering
 \begin{tabular}{  |c|c|c|c|c|c|}
 \hline
Topics& Paper &OTFS & MA Scheme &BS Antenna &  Characteristics\\ \hline
\multirow{6}*{OTFS-MA}& \cite{khammammetti2022spectral,habibi2024user}    & $\checkmark$        &OMA      & Single & \multirow{6}*{{\tabincell{l}{$\bullet$ Single-domain resource allocatioin\\ $\bullet$ Fixed MA scheme}}} \\
\cline{2-5}
~& \cite{ding2019otfs,ge2021otfs}                    & $\checkmark$        &power-domain NOMA     & Single     & ~  \\\cline{2-5}
~&\cite{ge2023otfs}                                & $\checkmark$        &SCMA     & Single     & ~  \\ \cline{2-5}
~&  \cite{ma2021otfs}                          & $\checkmark$        &TSMA     & Single     & ~  \\ \cline{2-5}
~& \cite{li2020new}                                 & $\checkmark$        &PDMA     & Multiple   & ~  \\ \cline{2-5}
~&  \cite{liu2023predictive}                               & $\checkmark$        &SDMA     & Multiple   & ~  \\ \hline
\multirow{2}*{MDMA}& \cite{liu2020multi,liu2020situation} & $\texttimes$     &{OMA, power/spatial-domain NOMA}   & Multiple   & \multirow{2}*{$\bullet$ Adaptive MA for stationary  channels} \\ \cline{2-5}
~& \cite{mei2022multi} & $\texttimes$     &{OMA, power/spatial/hybrid-domain NOMA}  & Multiple   & ~  \\ \hline
\tabincell{c}{OTFS-\\MDMA}& Our                                         & $\checkmark$        &{OMA/NOMA/SDMA}     & Multiple   & {\tabincell{l}{$\bullet$  Elastic multi-domain  resource allocation \\ $\bullet$  Adaptive MA  for mobility channels}}   \\ \hline
 \end{tabular}
 \renewcommand{\arraystretch}{1}
 }\vspace{-0.4cm}
\end{table*}

Nonetheless, the aforementioned studies exclusively concentrate on a single-domain or fixed access scheme, thereby exhibiting inherent limitations in terms of enhancing spectral efficiency.
As a remedy, the multi-dimensional MA (MDMA) method may be regarded as a promising access technique\cite{wang2023realizing}.
Explicitly, MDMA is a hybrid MA technology capable of flexibly apportioning interference levels across various radio resource domains by beneficially integrating different types of MA methods  (orthogonal or non-orthogonal).
Thus flexible channel access is capable of increasing the number of users while supporting diverse high quality-of-services (QoSs).
Specifically, the authors of \cite{liu2020multi},\cite{liu2020situation} explored how the non-orthogonality introduced by NOMA and SDMA affects the resource utilization costs in various radio resource domains.
They dynamically switched between OMA, power-domain NOMA, and spatial-domain NOMA to maximize the cost-aware system throughput. Then, these solutions were extended in \cite{mei2022multi}, where a hybrid spatial- and power-domain NOMA was proposed to glean additional resource diversity for performance enhancement.
They carefully considered the user-specific resource availability and the resource exploitation capability to strike a balance between the resource utilization cost and the performance gain.
However, these studies only address MDMA designs for stationary wireless channels.
Hence, their solutions suffer from significant performance degradation in high-mobility scenarios of future cellular systems.

Against the above backdrop, also summarized in Table \ref{table00},  we highlight our main contributions as follows:
\begin{itemize}
\item We develop an elastic multi-domain resource utilization mechanism for multi-user OTFS-MDMA systems, which leverages the time-invariant user-specific channel characteristics across the DD, power, and spatial domains to facilitate flexible channel access and resource allocation, thereby enabling efficient varied service provisions.
\item We conceive a novel DD domain division and user accommodation scheme, where the DD resource bins are partitioned into separate subregions called resource slots (RSs), and each RS is associated with a specific group of users based on their channel characteristics, thus reducing multi-user interference.
     This is termed as DD domain user accommodation.
     Then,  the most suitable MA techniques, including OMA/NOMA/SDMA, are applied for each RS based on their interference levels in the power and spatial domains, thus increasing the spectrum efficiency.
\item We jointly optimize the user accommodation, MA scheme selection, and power allocation for maximizing the weighted sum-rate of all users subject to individual minimum rate constraints and various practical constraints. Since the problem formulated belongs to the mixed integer nonlinear programming (MINLP) family, we develop an optimal algorithm based on dynamic programming and monotonic optimization (DPMO) for the problem in the special case of disregarding individual rate constraints.
Subsequently, we propose a low-complexity successive convex approximation-based simulated annealing (SCA-SA) algorithm to obtain a sub-optimal solution to the original problem in general cases.
\item Simulation results show that the proposed sub-optimal algorithm approaches the performance of the optimal algorithms, and significantly outperforms other baselines.
\end{itemize}

Organizations: Sections II and III introduce the OTFS-MDMA system model and problem formulation, respectively. Section IV develops the optimal solution for the relaxed problem and Section V develops the sub-optimal algorithm for the original problem. Finally, Section VI provides simulation results and Section VII concludes the paper.

Notations: The scalar, vector, and matrix are denoted by lowercase, bold lowercase, and bold uppercase letters, i.e.,  $x$, $\mathbf{x}$, and $\mathbf{X}$, respectively.
The transpose, conjugate transpose, diagonalization, Kronecker product,  vectorization, inverse vectorization, and Dirac delta function are denoted by $(\cdot)^T$, $(\cdot)^H$, ${\rm diag}(\cdot)$, $\otimes$, ${\rm vec}(\cdot)$, ${\rm vec}^{-1}(\cdot)$, and $\delta (\cdot)$, respectively.
Finally,  $\textstyle{{{\bf{F}}_X} = \left\{ {\frac{1}{{\sqrt X }}{e^{-{\rm j}2\pi \frac{{kl}}{X}}}} \right\}_{k,l = 0}^{X - 1}}$ represents the $X$-point discrete Fourier transform (DFT) matrix; ${\mathbb R}_{+}^{K}$ represents the set of nonegative real $K$-dimensional vectors; $\left| {\cal R} \right|$ denotes the number of elements in the set ${\cal R}$; and ${\bf 1}_R$ represents an $R$-dimensional vector with all elements equal to 1.

 \begin{figure*}[t]\centering
 \begin{minipage}{.48 \textwidth}
   \centering
   \includegraphics[width=\textwidth,height=3cm]{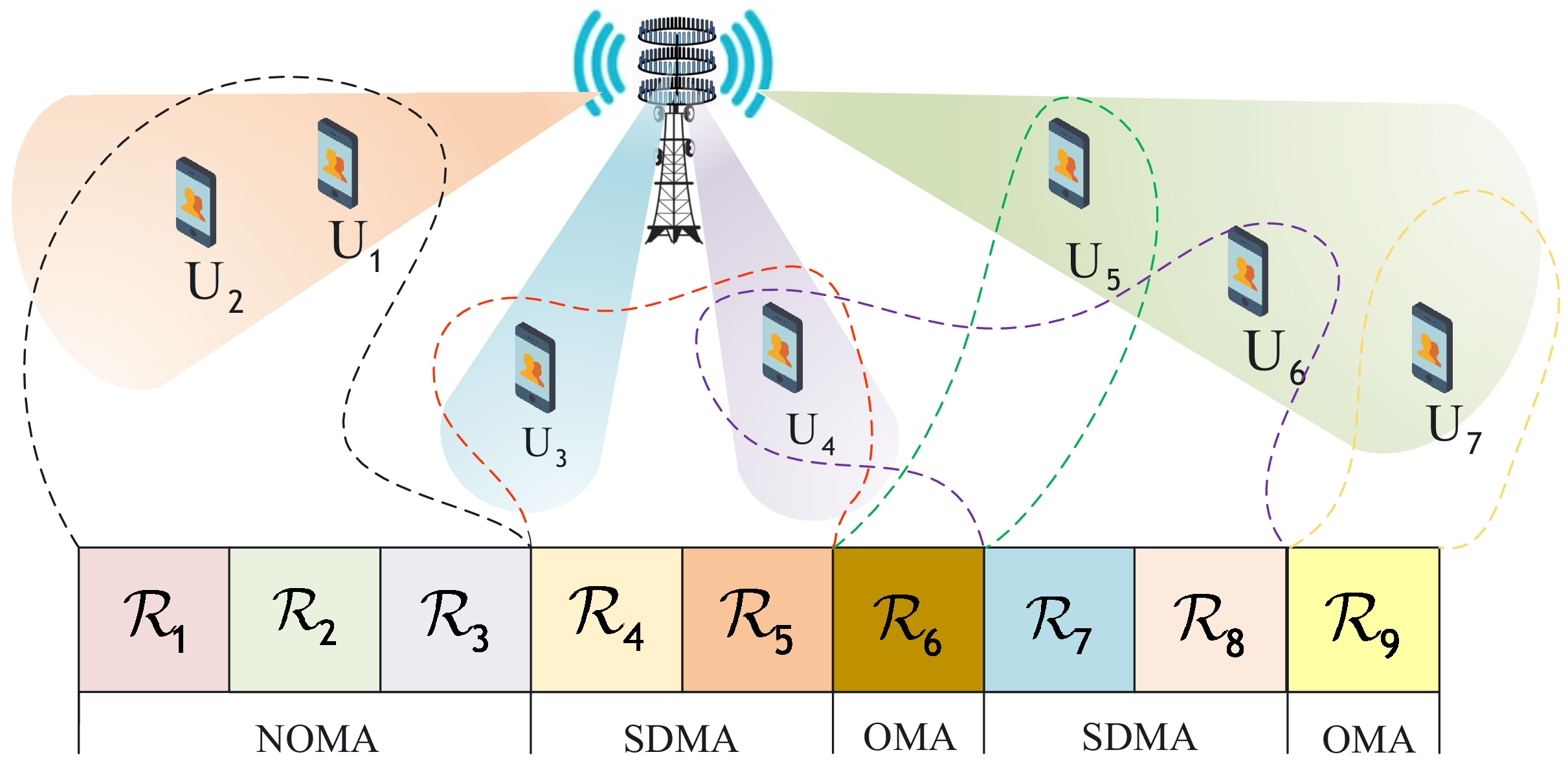}
   \subcaption{DD domain user accommodation by MDMA}\label{fig11}
  \end{minipage}
   \begin{minipage}{.4  \textwidth}
   \centering
   \includegraphics[width=\textwidth,height=3cm]{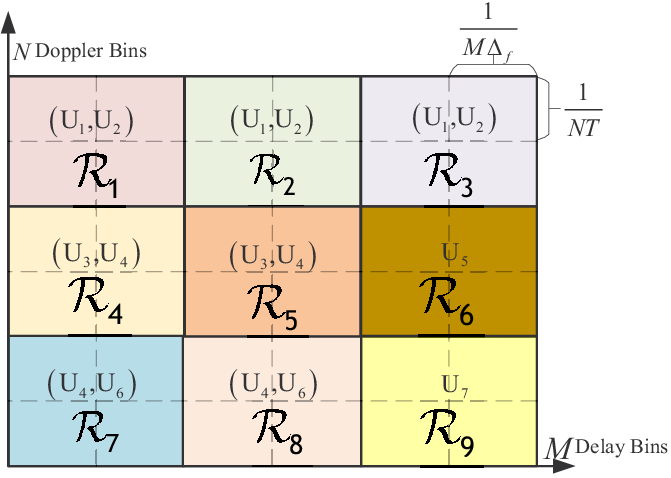}
   \subcaption{DD domain  division with $R=9$  resource slots (RSs), each of which includes
   ${\left| {{{{\cal R}}_r}} \right|}=\delta_M\delta_N=4$ resource bins}\label{fig12}
  \end{minipage}
   \caption{An illustration of user accommodation in the DD domain in a multi-user OTFS-MDMA system: $M=N=6$ and  $\delta_M=\delta_N=2$.
    }\label{fig1}
    \vspace{-0.3cm}
 \end{figure*}

\section{System Model}

As shown in Fig. \ref{fig1}, we consider a multi-user multiple-input single-output (MISO) OTFS-MDMA system, where a BS equipped with $D$ antennas broadcasts individual OTFS modulated signals to $Q$ single-antenna users, denoted by ${\rm U}_1,\cdots,{\rm U}_Q$.
We assume all channels are time-invariant in the DD domain during each transmission frame and the channel knowledge is perfectly known at the BS for the system design.

Next, by exploiting different degrees of orthogonality and interference of channels across the  DD, power, and spatial domains, we develop the OTFS-MDMA scheme.
Specifically, all DD resource bins are divided into multiple non-overlapping subregions called resource slots (RSs).
Taking into account the individual channel characteristics of each user in the multi-dimensional resource domains,
we assign specific groups of users to distinct RSs, which is called DD domain user accommodation.
Then, we apply MDMA to these RSs to achieve information transmission for these users.
This involves choosing the most suitable MA schemes from different resource domains, including NOMA and SDMA/OMA, tailored to each RS for enhancing spectral efficiency.
Intuitively,  the NOMA scheme is preferred by users equipped with powerful hardware for self-interference cancellation (SIC) and whose channels are highly correlated in the spatial domain, where the co-channel interference remains significant despite its mitigation by beamforming.
On the other hand, users having limited computational capabilities and orthogonal channels may adopt SDMA/OMA to achieve improved capacity at low complexity and interference.

\subsection{DD Domain User Accommodation }
In the DD domain, $MN$ data symbols are embedded in the two-dimensional $M\times N$ DD resource bins, where $M$ and $N$ denote the number of delay bins (or subcarriers) and the number of Doppler bins (or time intervals), respectively.
The subcarrier bandwidth and symbol duration are denoted by $\Delta_f$ and $T=1/\Delta_f$, respectively.
Let the $b$-th resource bin  represent the resource bin associated with the $(k+1)$-th Doppler bin (i.e., $\frac{k}{NT}$) and the $(l+1)$-th delay bin (i.e., $\frac{l}{M\Delta_f}$), where  $ {b}=kM+l+1$ for $0\le k\le N-1$ and $0 \le l \le M-1$.
As illustrated in Fig. \ref{fig1} (b), we assume that $M\times N$ DD resource bins are equally partitioned into $R$ RSs for simplicity, i.e., ${\cal R}_1,{\cal R}_2,\cdots, {\cal R}_R$, and each RS is characterized by lengths of $\delta_M$ and $\delta_N$.
 We assume that ${M}/{{{\delta _M}}}$ and ${N}/{{{\delta _N}}}$ are integers, and we know that $R=\frac{{MN}}{{{\delta _M}{\delta _N}}}$, ${\left| {{{{\cal R}}_r}} \right|}=\delta_M\delta_N$, $\bigcap {_{1 \le r \le R}{{\cal R}_r}}  = \emptyset $,  and $\bigcup\nolimits_{r = 1}^R {{ {\cal R}_r}}  = \left\{ {1,\cdots,b,\cdots,MN} \right\}$.

Next, a subset of users is accommodated in RS ${\cal R}_r$, where the choice of using NOMA or SDMA/OMA is based on the channel orthogonality among these users across the DD, power, and spatial domains.
Since increasing the number of users in the same RS increases both the interference and also the decoding complexity and delay in the NOMA scheme, we assume that each RS ${\cal R}_r$ only accommodates two users for NOMA \cite{sun2017optimal,chen2019optimal}, but the maximum of $Q$ users for SDMA.
Here, OMA can be regarded as a special case of SDMA if there is only a single user in an RS.

Then, let $a_{rqi}^{\rm N}$ and $a^{\rm S}_{r}$ denote binary variables involving user accommodation decisions and MA scheme selections, i.e.,  $a_{rqi}^{\rm N}=1$ means that users ${\rm U}_q$ and ${\rm U}_i$ are accommodated in ${\cal R}_r$ using NOMA, otherwise $a_{rqi}^{\rm N}=0$. Besides, $a_{r}^{\rm S}=1$ means that users are accommodated in ${\cal R}_r$ using SDMA/OMA, otherwise $a_{r}^{\rm S}=0$.
Then, let the data symbol of user ${\rm U}_q$ transmitted in the $b$-th resource bin be  $x_{qb}$.
Thus, 
 the $b$-th DD domain data symbol in RS ${{\cal R}_r}$ transmitted through the $d$-th antenna before multiplexing with the aid of the transmit precoding matrix is
\begin{align}
x_{{\rm{DD}}}^d\left[ b \right] = \left\{ {\begin{array}{*{20}{l}}
{\sqrt {{p_{qb}}} w_{qb}^d{x_{qb}} + \sqrt {{p_{ib}}} w_{ib}^d{x_{ib}},\;{\rm{if}}\;a_{rqi}^{\rm{N}} = 1},\\
{\sum\nolimits_{q = 1}^Q {\sqrt {{p_{qb}}} w_{qb}^d{x_{qb}}} ,\;{\rm{if}}\;a_r^{\rm{S}} = 1{\rm{ }},}
\end{array}} \right.
\end{align}
for $b \in {\cal R}_r$ and $\forall r$, where $w_{qb}^{d}$ is the $d$-th element of the transmit beamforming vector ${\bf w}_{qb}\in{\mathbb C}^{D\times1}$ with normalized power for the information transmission of ${\rm U}_q$ at the $b$-th DD resource bin. Besides,  $p_{qb}$ is the corresponding transmission power.

\subsection{OTFS Transmission Signal Model}
In this part, we introduce the OTFS downlink transmission signal model.
By rewriting $x_{{\rm{DD}}}^d\left[ b \right]$ for $1\le b\le MN$ into the vectorial form, i.e., ${{\bf{x}}^d_{\rm{DD}}}\in{\mathbb C}^{MN\times1}$ and denoting the transmit precoding matrix by ${\bf P}_{\rm tx}\in {{\mathbb C}^{MN\times MN}}$,  we have the following symbol matrix, i.e.,
\begin{align}
{{\bf{\tilde X}}_{{\rm{DD}}}^d} = {\rm{ve}}{{\rm{c}}^{ - 1}}\left( {{\bf P}_{\rm tx}{{\bf{x}}^d_{\rm{DD}}}} \right) \in {{\mathbb C}^{N\times M}}.\label{eq2}
\end{align}

By applying the inverse symplectic finite Fourier transform (SFFT), these data symbols in the TF domain transmitted through the $d$-th antenna can be represented by
\begin{align}
{S}^d_{\rm{TF}}\left[ {n,m} \right] = \frac{1}{\sqrt{MN}} \sum\limits_{k = 0}^{N - 1} {\sum\limits_{l = 0}^{M - 1} {{{ \tilde x}^d_{\rm{DD}}}\left[ {k,l} \right]{e^{{\rm{j}}2\pi \left( {\frac{{kn}}{N} - \frac{{ml}}{M}} \right)}}} },
\end{align}
where ${{ \tilde x}^d_{\rm{DD}}}\left[ {k,l} \right]$ is the $(k,l)$-th element of $
{{\bf{\tilde X}}_{{\rm{DD}}}^d}$.
Then, by applying the Heisenberg transform to the transmit waveform $G_{\rm tx}(t)$, the TF domain signals ${S}^d_{\rm{TF}}\left[ {n,m} \right] $ can be transformed into a continuous waveform in the time domain, yielding:
\begin{align}
&s^d \left( t \right) \nonumber\\
=&\sum\limits_{n = 0}^{N - 1} {\sum\limits_{m = 0}^{M - 1} {S^d_{\rm{TF}}\left[ {n,m} \right]} } {G_{\rm tx}}\left( t-nT \right){e^{{\rm j}2\pi m{\Delta _f}\left( {t - nT} \right)}}.
\end{align}

\subsection{Channel Model}
Let $h_q^{d}\left(\tau,\nu   \right)$ denote the channel response between the $d$-th antenna at the BS and user ${\rm U}_q$ in the DD domain.
This channel response comprises $L_q$ propagation paths, each with distinct delay $\tau _{q,\ell}$, Doppler shift $\upsilon _{q,\ell}$, and angle of departure  $\theta_{q,\ell}$, formulated as
\begin{align}
&{h_q^d}\left( {\tau ,\upsilon} \right) \nonumber\\
=&\sum\limits_{\ell= 1}^{{L_q}} {\frac{h_{q,\ell}}{\sqrt{L_q}} {{e^{ - {\rm j}\pi \sin {\theta _{q,\ell}}\left( {d - 1} \right)}}}\delta \left( {\tau  - {\tau _{q,\ell}}} \right)} \delta \left( {\upsilon  - {\upsilon _{q,\ell}}} \right),
\label{eqchannel}
\end{align}
where $h_{q,\ell}$ is the  complex channel response of the $\ell$-th propagation path arriving from the BS to user ${\rm U}_q$. Furthermore, ${\tau _{q,\ell}}$ and  ${\upsilon _{q,\ell}}$ can be represented by
${\tau _{q,\ell}} = \frac{{{l_{q,\ell}}}}{{M{\Delta _f}}}$ and ${\upsilon _{q,\ell}} = \frac{{{k_{q,\ell}} + {{\tilde k}_{q,\ell}}}}{{NT}}$, respectively, where ${l_{q,\ell}}$ and ${k_{q,\ell}}$ are integers representing the indices of ${\tau _{q,\ell}}$ and ${\upsilon _{q,\ell}}$, respectively. The real parameter $-0.5 \le{\tilde k_{q,\ell}}\le 0.5$ denotes the fraction of Doppler phase shift from the nearest integer ${k_{q,\ell}}$ associated with the $\ell$-th path. Here,  the fractional delay can be ignored due to the high-resolution sampling time $\frac{1}{M\Delta_f}$ \cite{yuan2021iterative}.

\subsection{OTFS Reception Signal Model}
From \eqref{eqchannel}, the  signal received in the time domain at user $q$ upon transmitting $\left\{s^d \left( t \right),1\le d\le D\right\}$ through $D$ antennas is formulated as:
\begin{align}
&y_q\left( t \right) \nonumber\\
=& \int{\!\int{\sum\limits_{d = 1}^D {h_q^d\left( {\tau ,\upsilon } \right){s^d}\left( {t \!-\! \tau } \right){e^{{\rm j}2\pi \upsilon \left( {t - \tau } \right)}}} } } {{\rm{d}}_\tau }{{\rm{d}}_\upsilon }+u_q\left(t\right),
\end{align}
where $u_q\left(t\right)$ is the complex additive white Gaussian noise.

Upon denoting the receive waveform by ${G_{\rm rx}}\left( {t }\right)$, the
signal received at ${\rm U}_q$ in the TF domain becomes\footnote{Applying additional TF windows on the transmitted signal ${S^d_{\rm{TF}}\left[ {n,m} \right]}$ and the received signal ${Y_{\rm TF}^q}\left[ {n,m} \right]$  can enhance channel estimation and data detection performance \cite{wei2021transmitter}. However, while this approach does not affect our system design and algorithm development, it can increase the complexity of signal processing. Therefore, we have chosen to omit it from this study.}:
\begin{align}
{Y_{\rm TF}^q}\left[ {n,m} \right] = \int {{y_q}} \left( t \right){G_{\rm rx}}\left( {t - nT} \right){e^{{\rm j}2\pi m{\Delta _f}\left( {t - nT} \right)}}{{\rm{d}}_t}.
\end{align}

Next, we assume that ${G}_{\rm tx}(t)$ and ${G}_{\rm rx}(t)$ satisfy the bi-orthogonal property, which can eliminate cross-symbol interference during symbol reception. This assumption provides an upper bound on the performance of OTFS with other waveforms, such as rectangular waveforms \cite{hadani2017orthogonal,raviteja2018interference}.
By applying the SFFT to ${Y_{\rm TF}^q}\left[ {n,m} \right] $, the channel's input-output relationship in the DD domain is expressed as \cite{raviteja2018interference}:
\begin{align}
y_{{\rm{DD}}}^q\left[ {k,l} \right]  =& \frac{1}{{\sqrt {MN} }}\sum\nolimits_{n = 0}^{N - 1} {\sum\nolimits_{m = 0}^{M - 1} {Y_{{\rm{TF}}}^q\left[ {n,m} \right]{e^{ - {\rm{j}}2\pi \left( {\frac{{nk}}{N} - \frac{{ml}}{M}} \right)}}} } \nonumber\\
 =& \frac{1}{{MN}}\sum\nolimits_{n = 0}^{N - 1} {\sum\nolimits_{m = 0}^{M - 1} {\left( {\sum\nolimits_{d = 1}^D {\tilde x_{{\rm{DD}}}^d\left[ {n,m} \right]} } \right.} } \nonumber\\
&\times\left. {\tilde h_q^d\left[ {{{\left( {k - n} \right)}_N},{{\left( {l - m} \right)}_M}} \right]} \right) + {u_q}\left[ {k,l} \right],
\end{align}
where ${u_q}\left[ {k,l}\right[ $ is the  noise term and
 $\tilde h_q^d\left( {k,l} \right) $ is given by
 \begin{align}
 \tilde h_q^d\left( {k,l} \right) = \sum\limits_{\ell  = 1}^{{L_q}} {h_{q,\ell }^d{e^{ - {\rm j}2\pi {\nu _{q,\ell }}{\tau _{q,\ell }}}}\phi \left( {k - {k_{q,\ell }},l - {l_{q,\ell }}} \right)},
\end{align}
with  $h_{q,\ell }^d = {h_{q,\ell }}{e^{ - {\rm j}\pi \sin {\theta _{q,\ell }}\left( {d - 1} \right)}}$ and $
\phi \left( {k - {k_{q,\ell }},l - {l_{q,\ell }}} \right) = \sum\nolimits_{n' = 0}^{N - 1} {\sum\nolimits_{m' = 0}^{M - 1} {{e^{ - {\rm{j}}2\pi \left( {n'\frac{{k - {k_{q,\ell }} - {{\tilde k}_{q,\ell }}}}{N} - m'\frac{{l - {l_{q,\ell }}}}{M}} \right)}}} }$.
Following a similar approximation of \cite{raviteja2018interference},  $y_{{\rm{DD}}}^q\left[ {k,l} \right]$ is rewritten by
\begin{align}
y_{{\rm{DD}}}^q\left[ {k,l} \right]\! \approx&\! \sum\limits_{d = 1}^D {\sum\limits_{\ell  = 1}^{{L_q}} {\sum\limits_{i =  - {N_{q,\ell }}}^{{N_{q,\ell }}} {\left( {\tilde x_{{\rm{DD}}}^d\left[ {{{\left( {k \!-\! {k_{q,\ell }} \!+ \!i} \right)}_N},{{\left( {l \!-\! {l_{q,\ell }}} \right)}_M}} \right]} \right.} } } \nonumber\\
&\times\left. {\bar h_{q,\ell ,i}^d{e^{ - {\rm j}2\pi {\nu _{q,\ell }}{\tau _{q,\ell }}}}} \right) + {u_q}\left[ {k,l} \right]
\label{eq14},
\end{align}
where $\bar h_{q,\ell ,i}^d = h_{q,\ell }^d\left( {\frac{{{e^{ - {\rm{j}}2\pi \left( { - i - {\tilde k_{q,\ell }}} \right)}} - 1}}{{N{e^{ - {\rm{j}}\frac{{2\pi }}{N}\left( { - i - {\tilde k_{q,\ell }}} \right)}} - N}}} \right)$. Here, $N_{q,\ell}\ll N$ is a small number, considering the significant values of $
\phi \left( {k - {k_{q,\ell }},l - {l_{q,\ell }}} \right) $ around  the peak at $k-k_{q,\ell}$.

Then, upon denoting the vectorial versions of $y_{{\rm{DD}}}^q\left[ {k,l} \right]$ and  $\tilde x_{{\rm{DD}}}^q\left[ {k,l} \right]$ by
${{\bf{y}}^q_{\rm DD}} \in{\mathbb C}^{MN\times1}$ and $
{{\bf{\tilde x}}^q_{\rm DD}} \in{\mathbb C}^{MN\times1}$, we rewrite \eqref{eq14} into the following matrix form  \cite{yuan2021iterative}
\begin{align}
{{\bf{y}}^q_{\rm DD}} = \sum\nolimits_{d = 1}^D {{\bf{H}}_q^d} {\bf{\tilde x}}^d_{{{\rm{DD}}}} + {{\bf{u}}^q_{\rm DD}}\label{eq15H},
\end{align}
where $ {{\bf{u}}^q_{\rm DD}} \in{\mathbb C}^{MN\times1}$ is the vectorial version of ${u_q}\left[ {k,l} \right]$, and ${{\bf{H}}_q^d}\in{\mathbb C}^{MN\times MN}$ is given by
\begin{align}
{\bf{H}}_q^d &= \sum\nolimits_{\ell  = 1}^{{L_q}} {\sum\nolimits_{i =  - {N_{q,\ell }}}^{{N_{q,\ell }}} {\left( {\bar h_{q,\ell ,i}^d{e^{ - {\rm j}2\pi \frac{{\left( {{k_{q,\ell }} + {{\tilde k}_{q,\ell }}} \right){l_{q,\ell }}}}{{MN}}}}} \right.} } \nonumber \\
& \qquad\times\left. {{{\bf{I}}_N}\left( { - {{\left[ {i - {k_{q,\ell }}} \right]}_N}} \right) \otimes {{\bf{I}}_M}\left( {{l_{q,\ell }}} \right)} \right)\label{eqhq}.
\end{align}
Here, ${{\bf{I}}_N}\left( { - {{\left[ {i - {k}} \right]}_N}} \right)$ is the matrix obtained upon  circularly shifting the rows of the $N \times N$ identity matrix by
$-[i-k]_N$, and ${{\bf{I}}_M}\left( {{l_{q,\ell }}} \right)\in{\mathbb C}^{M\times M}$ is defined similarly.
For the special case of no fractional Doppler phase shift, we have
\begin{align}{\bf{H}}_q^d = \sum\nolimits_{\ell  = 1}^{{L_q}} {h_{q,\ell }^d{e^{ - {\rm j}2\pi \frac{{ {{k_{q,\ell }}} {l_{q,\ell }}}}{{MN}}}}} {{\bf{I}}_N}\left( {{k_{q,\ell }}} \right) \otimes {{\bf{I}}_M}\left( {{l_{q,\ell }}} \right).
\end{align}

Note that regardless of whether there are fractional Doppler phase shifts, ${\bf H}_q^d$ is a block circulant matrix \cite{yuan2021iterative}, which can be formulated by ${\bf H}_q^d= {\rm{circ}}\left[ {{{\bf{G}}_{q,0}^d},{{\bf{G}}_{q,1}^d}, \cdots {{\bf{G}}_{q, N-1}^d}} \right]\in{\mathbb C}^{NM\times NM}$, i.e.,
\begin{align}
{\bf{H}}_q^d = \left[ {\begin{array}{*{20}{c}}
{{\bf{G}}_{q,0}^d}&{{\bf{G}}_{q,N - 1}^d}& \cdots &{{\bf{G}}_{q,1}^d}\\
 \vdots & \ddots & \ddots & \vdots \\
{{\bf{G}}_{q,N - 1}^d}& \cdots &{{{\bf{G}}_{q,1}}}&{{\bf{G}}_{q,0}^d}
\end{array}} \right],
\end{align}
where  ${{\mathbb{I}}_x}\left( y \right)$ is the indicator function, i.e., ${{\mathbb{I}}_x}\left( y \right)=1$ if $x=y$, and ${{\mathbb{I}}_x}\left( y \right)=0$ otherwise, and
\begin{align}{\bf{G}}_{q,n}^d &= \sum\nolimits_{\ell  = 1}^{{L_q}} {\sum\nolimits_{i =  - {N_{q,\ell }}}^{{N_{q,\ell }}} {\left( {{{\mathbb I}_n}\left( { - {{\left[ {i - {k_{q,\ell }}} \right]}_N}} \right)} \right.} } \nonumber\\
&\times\left. {\bar h_{q,\ell ,i}^d{e^{ - {\rm j}2\pi \frac{{\left( {{k_{q,\ell }} + {{\tilde k}_{q,\ell }}} \right){l_{q,\ell }}}}{{MN}}}}{{\bf{I}}_M}\left( {{l_{q,\ell }}} \right)} \right)\in{\mathbb C}^{M\times M}.\label{eqBqn}
\end{align}

\section{Problem Formulation: Weighted Sum-Rate Maximization }
In this section, we first derive the transmission rate within each RS using NOMA or SDMA/OMA for the studied OFTS-MDMA system.
Then, we formulate the weighted sum-rate maximization problem by jointly optimizing the multi-domain resource allocation and MA selection.

\subsection{Communication Performance Derivations}
This part derives the transmission rate of each user achieved in each RS under the given MA scheme.

To begin with, we note that the fractional Doppler shifts will cause ISI, as shown in \eqref{eq14}, which can be addressed by using zero-forcing (ZF) and MMSE linear schemes \cite{raviteja2019otfs}, message-passing schemes \cite{liu2021message}, and cross-domain schemes \cite{li2021cross,zhang2023cross}.
In this paper, we apply a similar approach to the ZF in \cite{raviteja2019otfs} and the unitary approximate message passing in \cite{yuan2021iterative} to address the ISI by utilizing the block circulant matrix characteristic of ${{\bf H}_q^d}$.
Specifically, upon denoting the $N$-point and $M$-point DFT matrices by ${{\bf{F}}_N}$ and ${{\bf{F}}_M}$, respectively, we can apply the receive precoding matrix ${\bf P}_{\rm rx}=\left( {{{\bf{F}}_N} \otimes {\bf{F}}_M^H} \right)$ on the signals in \eqref{eq15H}  and transmit precoding matrix of ${\bf P}_{\rm tx}=\left( {{\bf{F}}_N^H \otimes {{\bf{F}}_M}} \right)$ defined in \eqref{eq2}  to  express the effective channel's input-output relationship in the DD domain defined in \eqref{eq15H} as a sum of diagonal matrices, i.e.,
 \begin{align}
{{{\bf{\tilde y}}}_q} &= \left( {{{\bf{F}}_N} \otimes {\bf{F}}_M^H} \right){\bf{y}}_{{\rm{DD}}}^q \nonumber\\
&= \sum\nolimits_{d = 1}^D {{\bf{\Lambda }}_q^d{\bf{x}}_{{\rm{DD}}}^d}  + {{{\bf{\tilde u}}}_q},\label{eqdiag}
\end{align}
where ${\bm{\Lambda }}_q^d = {\rm{diag}}\left( {\sum\nolimits_{n = 0}^{N - 1} {{\bf{\Lambda }}_{q,n}^d{e^{ - {\rm{j2}}\pi \frac{{ni}}{{N}}}},0 \le i \le N - 1} } \right)$, ${\bm \Lambda} _{q,n}^d = {\rm{diag}}\left( {\sum\nolimits_{m = 0}^{M - 1} {g_{q,n}^d\left( {m,1} \right){e^{{\rm{j2}}\pi \frac{{mj}}{{M}}}}} ,0 \le j \le M - 1} \right)$, and $g_{q,n}^d\left( {m,1} \right)$ is the element located in the $m$-th row and the first column of the matrix ${\bf G}_{q,n}^d$, and ${{{\bf{\tilde u}}}_q}=\left( {{{\bf{F}}_N} \otimes {\bf{F}}_M^H} \right){{\bf{u}}^q_{\rm DD}}$.
For detailed derivations, please refer to Appendix \ref{theorem0proof}.

\subsubsection{Transmission Rate by NOMA}\label{eqdefini1}
Based on \eqref{eqdiag}, if the $r$-th RS ${\cal R}_r$ accommodates users ${\rm U}_q$ and ${\rm U}_i$  by NOMA, i.e., $a_{rqi}=1$, the received signal at user $k \in \left\{ {q,i} \right\}$ in DD resource bin $b$ is
\begin{align}
\label{eqdiag1}
{\rm{y}}_{kb}^{\rm{N}}& = \sum\limits_{d = 1}^D {h_{kb}^d\left( {\sqrt {{p_{qb}}} w_{qb}^d{x_{qb}} + \sqrt {{p_{ib}}} w_{ib}^d{x_{ib}}} \right) + {{\tilde u}_{kb}}}  \nonumber\\
& = {\bf{h}}_{kb}^H\left( {\sqrt {{p_{qb}}} {{\bf{w}}_{qb}}{x_{qb}} + \sqrt {{p_{ib}}} {{\bf{w}}_{ib}}{x_{ib}}} \right) + {{\tilde u}_{{kb}}},
\end{align}
for $b\in{{\cal R}_r}$, where the superscript ``$\rm N$'' indicates that the signals are related to the NOMA scheme; ${{\tilde u}_{kb}}$ is the equivalent Gaussian noise with mean zero  and power ${\varsigma _k}$.
Furthermore, we have ${\bf{w}}_{kb} =\left[{w_{kb}^1},\cdots,{w_{kb}^{D}}\right]^{T}\in{\mathbb C}^{D\times1}$ and ${\bf{h}}_{kb}=\left[h_{kb}^1 ,\cdots, h_{kb}^D \right]^{T}\in{\mathbb C}^{D\times 1}$, where $ h_{kb}^d$ is the $b$-th diagonal element of matrix ${\bf{\Lambda }}_k^d$ defined in \eqref{eqdiag}.

By using the popular maximum ratio transmission (MRT) beamforming for simplicity, we have $\textstyle{{{\bf{w}}_{qb}} = \frac{{{\bf{h}}_{qb}^H}}{{\left\| {{{\bf{h}}_{qb}}} \right\|}}}$.
Besides, we also assume that the power of the channel ${\bf{h}}_{qb}$ is larger than that of ${\bf{h}}_{ib}$ if $q< i$.
Then, in order to guarantee the successful SIC in NOMA \cite{khan2019joint,chen2019optimal}, the transmit power with the adopted beamforming should follow the condition that
${{\gamma _{qqb}}{p_{qb}}< {\gamma _{qib}}{p_{ib}}} $ if $q< i$, where we define $\textstyle{{\gamma _{qqb}} = {\left| {{\bf{h}}_{qb}^H{{\bf{w}}_{qb}}} \right|^2}}/\varsigma_q$ and $\textstyle{{\gamma _{qib}} = {\left| {{\bf{h}}_{qb}^H{{\bf{w}}_{ib}}} \right|^2}}/\varsigma_q$, respectively.
This implies that at user $q$, the effective power of the desired signal $x_{qb}$ is lower than the effective power of the interference signal $x_{ib}$. Consequently,
user $q$ can first decode the high-power signal $x_{ib}$ and then perform SIC to eliminate this interference before decoding its own low-power signal $x_{qb}$.
Conversely, at user $i$,  the effective power of desired signal $x_{ib}$ is higher than the effective power of the interference signal $x_{qb}$. Therefore, user $i$ decodes its own signal directly by regarding $x_{qb}$ as interference \cite{khan2019joint, chen2019optimal}.
Based on the above setup and assuming $a_{rqi}^{\rm N}=1$ and $q<i$,  the average transmission rates of user $q$ and user $i$ under NOMA in RS ${{\cal R}_r}$ are, respectively, given by\footnote{Please note that the inter-user interference still exists when applying MRT, but it decreases as the number of antennas increases \cite{chen2023impact}.}
\begin{subequations}\label{eqRNOMA002}
\begin{align}
C_{rqi}^{{\rm{N}},q}&= \frac{1}{{MN}}\sum\nolimits_{b \in {{\cal R}_r}} {{{\log }_2}\left( {1 + {\gamma _{qqb}}p_{qb}} \right)}\label{eqRNOMA1}, \\
C_{rqi}^{{\rm{N}},i}& = \frac{1}{{MN}}\sum\nolimits_{b \in {{\cal R}_r}} {{{\log }_2}\left( {1 + \frac{{{\gamma _{iib}}p_{ib}}}{{{\gamma _{iqb}}p_{qb} + 1}}} \right)}.\label{eqRNOMA2}
\end{align}\end{subequations}

\subsubsection{Transmission Rate by SDMA/OMA}
Based on \eqref{eqdiag}, if RS ${\cal R}_r$ accommodates multiple users by SDMA/OMA, i.e., $a_r^{\rm S}=1$, the corresponding  signals received at  ${\rm U}_q$ in the $b$-th DD resource bin  can be written as
\begin{align}
{\rm{y}}_{qb}^{\rm{S}} = {\bf{h}}_{qb}^H\sum\limits_{i = 1}^Q {\sqrt {{p_{ib}}} {{\bf{w}}_{ib}}{x_{ib}}}  + {{\tilde u}_{qb}},
\end{align}
for $b\in{{\cal R}_r}$, where the superscript ``$\rm S$'' indicates that the signals are related to the SDMA/OMA scheme.
Then, upon using MRT beamforming, the corresponding average transmission rate of ${\rm U}_q$ under SDMA in RS ${{\cal R}_r}$ is
\begin{align} \label{eqRsdma}
C_{rq}^{\rm{S}} = \frac{1}{{MN}}\sum\nolimits_{b \in {{\cal R}_r}} {{{\log }_2}\left( {1 + \frac{{{\gamma _{qqb}}{p_{qb}}}}{{\sum\nolimits_{i \ne q} {{\gamma _{qib}}{p_{ib}}}  + 1}}} \right)},
\end{align}
where ${\gamma _{qqb}}$ and ${\gamma _{qib}}$ are defined before  \eqref{eqRNOMA002}.

\subsubsection{Weighted Sum-Rate in Each RS}
Following \eqref{eqRNOMA002} and \eqref{eqRsdma}, the weighted sum-rate in RS ${\cal R}_r$ can be expressed as:
  \begin{align}
  {C_r} = \sum\limits_{q=1}^{Q-1} {\sum\limits_{i=q+1}^Q {a_{rqi}^{\rm{N}}\left( {{\alpha _q}C_{rqi}^{{\rm{N}},q} + {\alpha _i}C_{rqi}^{{\rm{N}},i}} \right)} }  + a_{r}^{\rm{S}}\sum\limits_{q = 1}^Q {{\alpha _q}C_{rq}^{\rm{S}}},
 \end{align}
where ${\alpha _q}$ are the constant priorities of users in the resource allocation, which can be related to their rate fairness objectives and computation capabilities \cite{chen2019optimal}.

\subsection{Weighted Sum-Rate Maximization}
In the multi-user MISO OTFS-MDMA system, we aim to jointly optimize the user accommodation, MA selection, and power allocation for maximizing the weighted sum-rate in all RSs.
For notational simplicity,  the power allocation is represented by the vector ${\bf{p}}\in{\mathbb C}^{QMN\times1}$, which includes all elements $ {{{p}}_{qb}} $ for each $ b  \in  {{\cal R}_r}$,  $\forall r$, and $\forall q$. The binary user accommodation variables for the SDMA scheme are denoted by the vector $ {\bf a}^{\rm S} \in{\mathbb C}^{R\times1}$, including all elements $ {a_{r}^{\rm{S}}}$ for $\forall r$.
Besides,  the binary user accommodation variables for the SDMA scheme are represented by the vector ${\bf{a}}^{\rm N}\in{\mathbb C}^{\frac{{Q\left( {Q - 1} \right)}}{2}R\times1} $, which consists of all elements $ {a_{rqi}^{\rm{N}}} $ for $\forall   r$  and user pairs, i.e., ${\cal Q}\buildrel \Delta \over = \left\{ {\left( {q,i} \right)\left| {1 \le q < i \le Q} \right.} \right\}$.
Then, the studied problem can be mathematically formulated as:
\begin{subequations}
\begin{align}\label{eqP1}
\mathop {\max }\limits_{{{\bf a}^{\rm N}},{{\bf a}^{\rm S}},{{\bf{p}} }} &\;\sum\limits_{r = 1}^R {{C_r}} \tag{{\bf{P1}}}\\
{\rm{s}}.{\rm{t}}.\;\;
&\sum\limits_{r = 1}^R {\left( {\sum\limits_{i = 1}^{q - 1} {a_{riq}^{\rm{N}}C_{riq}^{{\rm{N}},q}}  + \sum\limits_{i = q + 1}^Q {a_{rqi}^{\rm{N}}C_{rqi}^{{\rm{N}},q} + } a_{r}^{\rm{S}}C_{rq}^{\rm{S}}} \right)}  \nonumber\\
&\qquad\qquad\qquad\qquad\qquad\qquad\quad\ge {C_{\min ,q}},\forall q,\label{eqP1a0} \\
&\sum\limits_{r = 1}^R {\sum\limits_{b \in {{\cal R}_r}} { \left( {\sum\limits_{q = 1}^{Q-1} {\sum\limits_{i = q + 1}^Q {a_{rqi}^{\rm{N}}\left( {{p_{qb}} + {p_{ib}}} \right) + a_{r}^{\rm{S}}\sum\limits_{q = 1}^Q {{p_{qb}}} } } } \right)} }  \nonumber\\
&\qquad\qquad\qquad\qquad\qquad\qquad\qquad\qquad \le P, \label{eqP1a}\\
&a_{rqi}^{\rm{N}}\left( {{\gamma _{qqb}}{p_{qb}} - {\gamma _{qib}}{p_{ib}}} \right) \le 0,  \nonumber\\
&\qquad\qquad\qquad\qquad \forall b \in {{\cal R}_r},\forall r,\forall\left( {q,i} \right) \in {\cal Q},\label{eqP1a1}\\
&\sum\nolimits_{q = 1}^{Q-1}  {\sum\nolimits_{i = q + 1}^Q {a_{rqi}^{\rm{N}}} }  + a_r^{\rm{S}} = 1,\forall r,\label{eqP1b0}\\
& p_{qb}\ge0,  \forall b \in {{\cal R}_r},\forall r,\forall q,\label{eqP1d}\\
& a_{rqi}^{\rm{N}} \in \left\{ {0,1} \right\},a_r^{\rm{S}} \in \left\{ {0,1} \right\}, \forall r,\forall\left( {q,i} \right) \in {\cal Q}.\label{eqP1c}
\end{align}\end{subequations}
To elaborate, \eqref{eqP1a0} is the individual rate constraint, which implies that the minimal transmission rate of user ${\rm U}_q$ should be higher than a constant rate $C_{{\rm min},q}$; \eqref{eqP1a} represents the total power constraint;
\eqref{eqP1a1} is the power constraint determining the decoding order within the NOMA protocol \cite{zhu2017optimal,chen2019optimal}.
Furthermore, \eqref{eqP1b0} is the binary user accommodation and MA optimization constraint, which indicates that each RS is either assigned to NOMA or  SDMA/OMA, and  \eqref{eqP1d} and \eqref{eqP1c} represent additional practical constraints.

However, $\bf P1$ is a non-convex MINLP problem, which is generally NP-hard \cite{chen2020joint,chen2023wireless,jia2023new}. Hence, the exhaustive search is usually applied to find the optimal solution, which results in complexity issues. This motivates us to develop an efficient algorithm to solve this problem.

\section{Optimal Solution to The Special Case: Without Rate Constraints}
This section proposes an efficient dynamic programming and monotonic optimization (DPMO) algorithm to find the optimal solution in the special case of disregarding the individual rate constraints, i.e., ${C_{\min,q}}=0$.
Specifically, we first transform problem \ref{eqP1} into a DP recursion framework. Then, since each recursion has to solve a non-convex sub-problem, we transform it into a MO form and apply the  branch-reduction-and-bound (BRB) algorithm to find the optimal solution.

\subsection{Dynamic Programming Recursion Framework}
We find that problem  \ref{eqP1} can be partitioned into $R$ independent problems if ${C_{\min,q}}=0$ and the power allocated to RS ${\cal R}_r$ is already given. Thus, to find the optimal solution of \ref{eqP1}, we transform it into a DP recursion framework by defining the resource allocation state at the $r$-th RS ${\cal R}_r$ by
\begin{align}
E_r=\sum\limits_{i = 1}^r {{p_i}},\; {\rm for}\;1\le r\le R,
 \end{align}
where ${p_i}$ is the power allocated to RS  ${\cal R}_i$. Here,  the resource allocation state $E_r$ in the $r$-th RS denotes the total power allocated to the first $r$ RSs. Then, it is straightforward to know that $E_0=0$. Similarly, following the analysis in \cite{chen2019optimal}, we know that $E_R=P$ since a higher transmission power achieves a higher transmission rate.

Then, the state transition probability among $R$ RSs can be modeled by a Markov chain, which means that the remaining amount of power that can be optimized and allocated to RS ${\cal R}_r$ only depends on the previous state $E_{r-1}$.
Governed by the Markov state transition, the weighted sum-rate achieved in the first $r$ DD RSs is rewritten as:
\begin{align}
\sum\limits_{i = 1}^r {{C_r}}  = \sum\limits_{i = 1}^{r-1} {{C_r}}  + \Delta C\left( {{E_{r - 1}},{E_r}} \right),
\end{align}
where $\Delta C\left( {{E_{r - 1}},{E_r}} \right)$ denotes the weighted sum-rate achieved in  RS  ${\cal R}_r$, when the state changes from $E_{r - 1}$ to $E_{r }$.

Based on the above definitions, the optimal solution to problem  \ref{eqP1} can be obtained through the following DP recursion framework:
\begin{subequations}\label{eqDp}
\begin{align}
{C^\star}\left( {{E_r}} \right) &= \mathop {\max }\limits_{\forall {E_{r - 1}}} \left\{ {{\Delta ^\star}C\left( {{E_{r - 1}},{E_r}} \right)+{C^\star}\left( {{E_{r - 1}}} \right)} \right\},\nonumber\\
&\quad\qquad\qquad\qquad\qquad\qquad\;\;{\rm for}\; 2\le r\le R,\\
{C^\star}\left( {{E_1}} \right) &= {\Delta ^\star}C\left( {{E_{0}},{E_1}} \right),
\end{align}\end{subequations}
where ${C^\star}\left( {{E_r}} \right) $ is the optimal weighted sum-rate achieved in the $r$-th RS when the state goes to $E_r$.
Furthermore,  ${{\Delta ^\star}C\left( {{E_{r - 1}},{E_r}} \right)}$ is the optimal $C_r$,  when the state changes from $E_{r - 1}$ to $E_{r }$. Moreover, ${{\Delta ^\star}C\left( {{E_{r - 1}},{E_r}} \right)}$ can be mathematically obtained from
\begin{align}\label{optp2}
{\Delta ^\star}C\left( {{E_{r - 1}},{E_r}} \right)  \buildrel \Delta \over =  \max \left\{ {\mathop {\max }\limits_{\forall a_{rqi}^{\rm{N}}} C_{rqi}^{{\rm{N}}, \star },C_r^{{\rm{S}}, \star }} \right\} \tag{\bf P2}.
\end{align}
Here,  $C_{rqi}^{{\rm{N,\star}}}$ is derived from the following problem  by setting $a_{rqi}^{\rm N}=1$ and $a_{r}^{\rm S}=0$ under NOMA, i.e.,
\begin{align}\label{sp2a}
C_{rqi}^{{\rm{N}}, \star } \buildrel \Delta \over = \mathop {\max }\limits_{{{\bf{p}}}} &  \sum\limits_{b \in {{\cal R}_r}} {\left( {{{\tilde \alpha }_q}{{\log }_2}\left( {1 + {\gamma _{qqb}}{p_{qb}}} \right)} \right.} \nonumber\\
&\qquad\quad\left. { + {{\tilde \alpha }_i}{{\log }_2}\left( {1 + \frac{{{\gamma _{iib}}{p_{ib}}}}{{{\gamma _{iqb}}{p_{qb}} + 1}}} \right)} \right)
 \tag{\bf P2a}\\
{\rm{s}}.{\rm{t}}.&\; \sum\limits_{b \in {{\cal R}_r}} {\left( {{p_{qb}} + {p_{ib}}} \right)}  \le {p_r},
\eqref{eqP1a1}, \;{\rm and}\; \eqref{eqP1d},\label{eq31}
\end{align}
where ${\tilde \alpha }_q=\frac{{ \alpha }_q}{MN}$ for $1\le q\le Q$, and $p_r=E_r-E_{r-1}$ denotes the total power allocated to RS ${\cal R}_r$. Similarly, ${C_r^{{\rm{S,\star}}}}$ is obtained from solving the following problem by setting $a_{r}^{\rm S}=1$ and $ {a_{rqi}^{\rm{N}}}  = 0$ under SDMA/OMA, i.e.,
\begin{align}\label{sp2b}
\!\!\!\!\!\!C_r^{{\rm{S}}, \star }  \buildrel \Delta \over =  \mathop {\max }\limits_{\bf{p}}& \sum\limits_{b \in {{\cal R}_r}} {\sum\limits_{q = 1}^Q {{{\tilde \alpha }_q}{{\log }_2}\left( {1 + \frac{{{\gamma _{qqb}}{p_{qb}}}}{{\sum\nolimits_{i \ne q}^Q {{\gamma _{qib}}{p_{ib}}}  + 1}}} \right)} } \tag{\bf P2b}\\
{\rm{s}}.{\rm{t}}.& \sum\limits_{b \in {{\cal R}_r}} {\sum\limits_{q = 1}^Q {{p_{qb}}} }  \le {p_r},
 \;{\rm and}\; \eqref{eqP1d}.\label{eq32}
\end{align}
To solve problem  \ref{optp2}, we have to solve problem \ref{sp2a}  $\frac{{Q\left( {Q - 1} \right)}}{2}$ times because there are  $\frac{{Q\left( {Q - 1} \right)}}{2}$  different combinations of $a_{rqi}^{\rm N}$ in RS ${\cal R}_r$, and further solve problem  \ref{sp2b} once. Finally, if problem \ref{optp2} can be solved optimally, the optimal solution to problem \ref{eqP1}  can be obtained by using the DP framework in \eqref{eqDp}, which is summarized in Algorithm \ref{algorithmDPMO1}.

However, obtaining optimal solutions to problems \ref{sp2a} and \ref{sp2b} is challenging due to the associated non-convex fractional terms.
While the widely used fractional programming \cite{chen2024learning} can efficiently solve both problems, it can only guarantee stationary solutions instead of the optimal solution.
This limitation diminishes the significance of our efforts in proposing the DP framework since we lose the optimality of the original problem \ref{eqP1}.
Therefore, we introduce the following MO technique to find the optimal solutions of \ref{sp2a} and \ref{sp2b}.

\begin{algorithm}[t]
\caption{Optimal solution (DPMO) to problem $\bf P1$}\label{algorithmDPMO1}
\renewcommand{\baselinestretch}{0.9}
 {{
\begin{algorithmic}[1]
 \STATE Initialize $E_0=0$, $E_R=P$, ${{  C}^*}({E_{0}})=0 $, and quantize $P$ as $[0:\frac{P}{{\Delta _p}}:P]$ with resolution accuracy $\frac{P}{{\Delta _p}}$
	\FOR{ $r=1, \dots, R$}
        \FOR{power $p_r=[0:\frac{P}{{\Delta _p}}:P]$, and all possible $a_{rqi}^{\rm N}$ and $a_r^{\rm S}$ on RS ${\cal R}_r$ }
        \STATE Calculate and save $C_{rqi}^{{\rm{N}}, \star }$ by solving  \ref{sp2a}, i.e.,
        using Algorithm 2 to solve the equivalent \ref{p2aMO}
        \STATE Calculate and save  $C_r^{{\rm{S}}, \star }$  by solving \ref{sp2b}, i.e.,
          using Algorithm 2 to solve the equivalent  \ref{p2bMO}
        \ENDFOR
         \FOR{each state $E_r=[0:\frac{P}{{\Delta _p}}:P]$ on RS ${\cal R}_r$ }
         \FOR {all possible state $E_{r-1}=[0:\frac{P}{{\Delta _p}}:E_r]$ on RS ${\cal R}_{r-1}$ }
         \STATE Set $p_r=E_r-E_{r-1}$ and calculate $\Delta\!^* { C}( {{E_{r \!-\! 1}},{E_r}} )$ in problem \ref{optp2} by using results obtained in steps 4 and 5
             \STATE Calculate and save  ${{\cal D}}({\!E_{r\!-\!1}}\!)=\!\Delta\!^* { C}( {{E_{r \!-\! 1}},{E_r}} )\! +\! {{  C}^*}({E_{r \!- \!1}})$
         \ENDFOR
           \STATE Calculate ${{C}^*}({{E}_r}) = \mathop {\max }\nolimits_{\forall {E_{r - 1}}} {\cal D}({E_{r-1}}) $
         \ENDFOR
     \ENDFOR
\STATE Recover the optimal  ${\bf{a}^{\rm N}}$, ${\bf{a}^{\rm S}}$, and  ${{\bf{p}} }$ by backtracking on the state transition path, one by one, from $E_R$ to $E_0$ that achieves maximum ${{C}^*}({E_R})$.
\end{algorithmic}
}}
\end{algorithm}

\subsection{Preliminaries of MO and  Problem Transformation}\label{MOA}

In this part, we first introduce the basic concepts in MO \cite{zhang2013monotonic}. Subsequently, we reformulate problems  \ref{sp2a} and \ref{sp2b} as MO forms, allowing us to employ standard MO algorithms to obtain their optimal solutions.
\begin{itemize}
\item {\bf Box}: Given any $K$-dimensional vectors ${\bf b}\in{\mathbb R}_{+}^{K}$ and $ {\bf c}\in{\mathbb R}_{+}^{K}$ associated with $\bf b\le c$, a box with lower vertex $\bf b$ and upper vertex $\bf c$  is defined by the hyper rectangle $\left[ {{\bf{b}},{\bf{c}}} \right] = \left\{ {{\bf{z}}\left| {{\bf{b}} \le {\bf{z}} \le {\bf{c}}} \right.} \right\}$.
\item {\bf Normal}: An infinite set ${\cal Z} \subset {\mathbb R}_{+}^{K} $ is said to be normal if the box follows  $\left[ {{\bf{0}},{\bf{z}}} \right]\subset {\cal Z}$ for any element ${\bf z}\in{\cal Z}$.
\item {\bf MO problem}: An optimization problem belongs to the MO family if it can be written as
   \begin{align}
\mathop {\max }\limits_{\bf{z}} {\cal L}\left( {\bf{z}} \right)\quad {
\rm s.t.}\; {\bf{z}} \in {\cal Z}, \label{MO}
\end{align}
where $ {\cal L}\left( {\bf{z}} \right)$  is monotonically increasing with respect to $\bf z$ on ${\mathbb R}_{+}^{K}$ and $ {\cal Z}$ is a normal closed set. The globally optimal solution of problem \eqref{MO} can be obtained using the BRB algorithm \cite{bjornson2012robust}.
\end{itemize}

\subsubsection{Transform \ref{sp2a} to an MO Problem}
By introducing the auxiliary optimization vector ${\bf{z}}_r^{\rm N}= {\left[ {z_{qib}^{{\rm N},q},z_{qib}^{{\rm N},i}\left| {\forall b \in {{\cal R}_r}} \right.} \right]^T} \in {{\mathbb R}_{+}^{2{\left| {{{\cal R}_r}} \right|}}}$, problem \ref{sp2a} can be equivalently transformed into the following MO form, i.e.,
\begin{align}\label{p2aMO}\tag{\bf P2a-MO}
\mathop {\max }\limits_{{\bf{z}}_r^{\rm{N}}}\;   {\cal L}^{\rm N}\left( {{{\bf z}_r^{\rm N}}}\right)
 \quad {\rm{s}}{\rm{.t}}{\rm{.}} \;{\bf{z}}_r^{\rm{N}} \in {\cal Z}_r^{\rm{N}},
\end{align}
where
$ {\cal L}^{\rm N}\left( { {{\bf z}_r^{\rm N}}}\right) \buildrel \Delta \over =\sum\nolimits_{b \in {{\cal R}_r}} {\left( {{{\tilde \alpha }_q}z_{qib}^{{\rm N},q} + {{\tilde \alpha }_i}z_{qib}^{{\rm N},i}} \right)}  $. Here, ${\cal Z}_r^{\rm N}$ is the feasible region of ${\bf{z}}_r^{\rm N}$, i.e.,
\begin{align}
{\cal Z}_r^{\rm{N}} \!=\!\left\{ {{\bf{z}}_r^{\rm{N}}\left| {z_{qib}^{{\rm N},q} \le c_{qib}^{{\rm N},q}\left( {\bf{p}} \right),z_{qib}^{{\rm N},i} \le c_{qib}^{{\rm N},i}\left( {\bf{p}} \right),\forall b \in {{\cal R}_r}, \eqref{eq31}} \right.} \right\},
\end{align}
with $c_{qib}^{{\rm N},q}\left( {\bf{p}} \right) = {\log _2}\left( {1 + {\gamma _{qqb}}{p_{qb}}} \right)$ and $c_{qib}^{{\rm N},i}\left( {\bf{p}} \right) = {\log _2}\left( {1 + \frac{{{\gamma _{iib}}{p_{ib}}}}{{ {{\gamma _{iqb}}{p_{qb}} + 1} }}} \right)$.

\subsubsection{Transform  \ref{sp2b} to an MO Problem}
By introducing the optimization vector ${\bf{z}}_r^{\rm S} = {\left[ {z_{qb}^{\rm S}\left| {\forall b \in {{\cal R}_r}},\forall q \right.} \right]^T} \in {{\mathbb R}_{+}^{Q{\left| {{{\cal R}_r}} \right|}}}$, problem \ref{sp2b} becomes equivalent to the following MO form:
\begin{align}\label{p2bMO}
\mathop {\max }\limits_{{\bf z}_r^{\rm{S}}}\;{\cal L}^{\rm S}\left( {{{\bf z}_r^{\rm S}}}\right) \quad {\rm{s}}{\rm{.t}}{\rm{.}}\;{\bf{z}}_r^{\rm{S}} \in {\cal Z}_r^{\rm{S}},\tag{\bf P2b-MO}
\end{align}
where ${\cal L}^{\rm S}\left( { {{\bf z}_r^{\rm S}}}\right) \buildrel \Delta \over = \sum\nolimits_{ b \in {{\cal R}_r}} {\sum\nolimits_{q = 1}^Q {{{\tilde \alpha }_q}z_{qb}^{\rm{S}}} } $. Here, ${\cal Z}_r^{\rm S}$ is the feasible set of ${\bf{z}}_r^{\rm S}$, i.e.,
\begin{align}
{\cal Z}_r^{\rm{S}}  = \left\{ {{\bf{z}}_r^{\rm{S}}\left| {z_{qb}^{\rm{S}} \le c_{qb}^{\rm{S}}\left( {\bf{p}} \right),\forall b \in {{\cal R}_r},\forall q, \eqref{eq32}} \right.} \right\},
\end{align}with $c_{qb}^{\rm{S}}\left( {\bf{p}} \right) = {\log _2}\left( {1 + \frac{{{\gamma _{qqb}}{p_{qb}}}}{{\sum\nolimits_{i \ne q}^Q {{\gamma _{qib}}{p_{ib}}}  + 1}}} \right)$.

\begin{figure}[ht]\centering
 \begin{minipage}{.5  \textwidth}
   \centering
   \includegraphics[width=.9\textwidth, height=3.4cm]{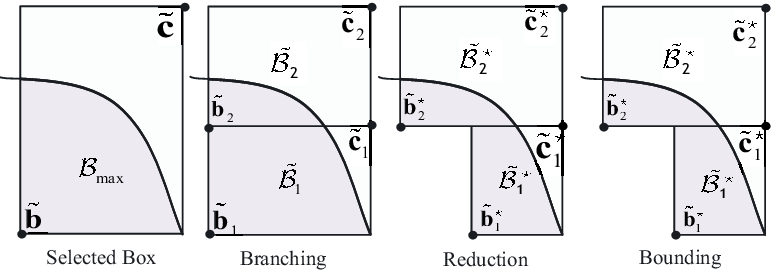}
   \subcaption{Steps in one iteration}\label{figsym22}
  \end{minipage}
   \begin{minipage}{.48  \textwidth}
   \centering
   \includegraphics[width=.95\textwidth, height=6.4cm]{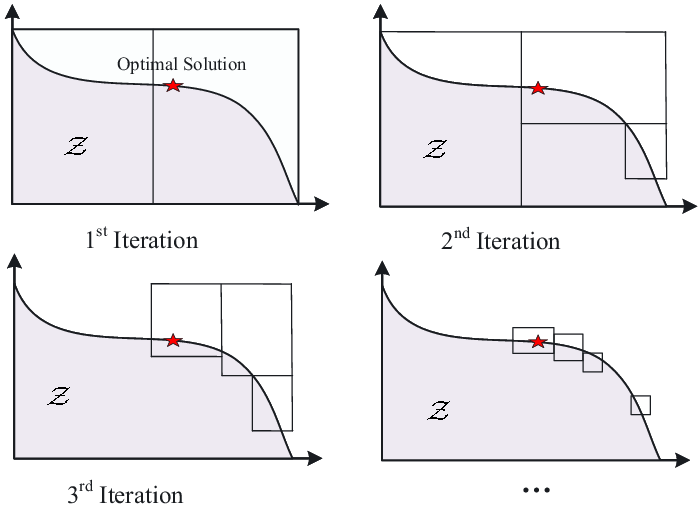}
   \subcaption{Box refinement in each iteration}\label{figsym23}
  \end{minipage}
  \caption{Illustration of the BRB for MO problem: It determines the optimal solution by approximating the feasible region with a set of boxes. The box set is refined iteratively, with regions that cannot contain the optimal point removed in each iteration.}\label{BRBillustration}
\end{figure}

\subsection{BRB Algorithm for MO}
In this part, we first present the principle of the BRB algorithm to find optimal solutions to the general MO problems in \eqref{MO}.
Subsequently, we utilize the BRB algorithm to solve \ref{p2aMO} and \ref{p2bMO} and illustrate the details using the pseudocode in Algorithm \ref{algorithm2}.
To begin with, we briefly introduce the fundamental principles of solving the general MO problem defined in \eqref{MO}, also illustrated in Fig. \ref{BRBillustration}:
\begin{itemize}
\item Since ${\cal L}\left( {\bf{z}} \right)$ is monotonically increasing with respect to $\bf z$, the optimal solution must lie on the boundary of the feasible region $\cal Z$.
To find the optimal solution, we use a box set $\cal N$, consisting of multiple non-overlapping boxes, to approximate the boundary of the feasible region $\cal Z$ that contains the optimal solution.

\item For one box ${\cal B}=\left[ {{\bf{b}},{\bf{c}}} \right]$ belonging to the box set $\cal N$, i.e., ${\cal B} \in {\cal N}$,
we denote the lower and upper bounds of all points in ${\cal B}$ by ${f^{\rm lb}\left(\cal B\right)}$ and ${f^{\rm ub}\left(\cal B\right)}$.
Thanks to the monotonically increasing objective function, we know ${f^{\rm lb}\left(\cal B\right)}={{\cal L}\left( {\bf{b}} \right)}$ and $f^{\rm ub}\left(\cal B\right)={{\cal L}\left( {\bf{c}} \right)}$.

\item The maximum lower and upper bounds can be found by comparing the lower and upper bounds of all boxes of $\cal N$, and we denote them by $f^{\rm min}_{\cal N}$ and  $f^{\rm max}_{\cal N}$, respectively.
  Then, we iteratively split certain boxes into sub-boxes and try to improve $f^{\rm min}_{\cal N}$ and reduce $f^{\rm max}_{\cal N}$. The algorithm will converge to global optimality with a predefined positive accuracy parameter $\varepsilon$, i.e., $f^{\rm max}_{\cal N}<(1+\varepsilon)f^{\rm min}_{\cal N}$.
\end{itemize}

Next, we introduce the detailed procedures of the BRB algorithm for solving MO problems. Firstly, we establish an initial box set ${\cal N}=\left\{{\cal B}_0\right\}$ where ${\cal B}_0=[0,{\bf z}_0]$ is the box that contains the whole feasible set ${\cal Z} \subset {\mathbb R}_{+}^{K} $. We initialize the lower and upper bounds  by $f^{\rm min}_{\cal N} = {\cal L}\left( {\bf{0}} \right) = 0$  and $f^{\rm max}_{\cal N}= {\cal L}\left( {\bf{z_0}} \right)$.
Then, each iteration of the BRB consists of three steps, i.e., branching, reduction, and bounding.
\subsubsection{\bf Branching} We first find the specific box ${{\cal B}_{\max }} $ from the box set $\cal N$ that achieves the upper bound $f^{\rm max}_{\cal N}$, i.e.,
 \begin{align}
{{\cal B}_{\max }} \buildrel \Delta \over =  \mathop {\arg \max }\nolimits_{{\cal B} \in {\cal N}} {f^{{\rm{ub}}}}\left( {\cal B} \right). \label{eq31a}
\end{align}
Then, upon denoting ${{\cal B}_{\max }}=\left[ {{\tilde {\bf{b}}},{\tilde{\bf{c}} }} \right]$, we bisect ${{\cal B}_{\max }}$ along its longest side, resulting in two new sub-boxes, i.e.,
\begin{align}
{\tilde{\cal  B}_1} = \left[ {{\tilde{\bf{b}}},{\tilde{\bf{c}}} - \xi {{\bf{e}}_{\dim }}} \right],{\tilde{\cal B}_2} = \left[ {{\tilde{\bf{b}}} + \xi {{\bf{e}}_{\dim }},{\tilde{\bf{c}}}} \right],\label{eq39}
\end{align}
where $\dim  = \mathop {\arg \max }\nolimits_i \left\{ {{{\tilde c}_i} - {{\tilde b}_i}} \right\}$ with  ${\tilde c}_i$ and ${\tilde b}_i$ being the $i$-th element of $\tilde{\bf c}$ and $\tilde{\bf b}$, respectively. Besides,   $\xi  = \frac{{{{\tilde c}_{\dim }} - {{\tilde b}_{\dim }}}}{2}$ and ${\bf e}_i$ is the $i$-th column of ${\bf I}_{K}$. Then, the upper bounds of each box are given by
\begin{subequations}\label{eq40}
\begin{align}
{f^{{\rm{ub}}}}\left( {{{\tilde {\cal B}}_1}} \right) &= \min \left\{ {{f^{{\rm{ub}}}}\left( {\cal B} \right),{\cal L}\left( {\tilde{\bf{c}}} - \xi {{\bf{e}}_{\dim }} \right)} \right\},\\
{f^{{\rm{ub}}}}\left( {{{\tilde {\cal B}}_2}} \right) &={\cal L}\left( {\tilde{\bf{c}}} \right).
\end{align}
\end{subequations}
\subsubsection{\bf Reduction}
This step removes points of the  two sub-boxes that cannot be the optimal solution.
Specifically, from \eqref{eq39}, we denote the divided
sub-boxes by
${\tilde{\cal  B}_l}=\left[ {{\bf{\tilde b}}_l,{\bf{\tilde c}}}_l \right]  $ for $l=1, 2$ with each corresponding reduced box represented by ${\tilde {\cal B}}_l^\star  \buildrel \Delta \over = \left[ {{\bf{\tilde b}}_l^\star,{\bf{\tilde c}}}_l^\star \right]$.
Then, by exploiting monotonicity, we know that ${\tilde{\cal B}_l}$ cannot contain the optimal solution and can be removed, i.e., ${\tilde {\cal B}}_l^\star =\emptyset $,  if it meets any of the following constraints:
\begin{itemize}
\item ${\bf{\tilde b}}_l$ is not in the feasible region: because this implies that all vertices $\bf z$ belonging to box ${\tilde{\cal B}_l}$ satisfy ${\bf{\tilde b}}_l \le \bf z$ and thus lie outside the feasible region;
\item  ${f^{{\rm{ub}}}}\left( {\tilde{\cal  B}_l}  \right)\le f^{\rm min}_{\cal N}$: because this implies that there already exists a feasible solution that is larger than  ${\cal L}({\bf z})$ for any vertex $\bf z$  in ${\tilde{\cal  B}_l}$.
\end{itemize}

On the other hand,  if ${f^{{\rm{ub}}}}\left( {\tilde{\cal  B}_l}  \right)\ge f^{\rm min}_{\cal N}$,  we can also remove redundant points of ${\tilde{\cal  B}_l}$  that cannot be the optimal solution, as illustrated in Fig. \ref{figsym22}.
Specifically, from \cite{bjornson2012robust}, we know that all points ${\bf x}\in{{\tilde {\cal B}}_l}$ satisfying $f^{\rm min}_{\cal N}\le {\cal L}\left(\bf x\right)\le {f^{{\rm{ub}}}}\left( \tilde{\cal B}_l \right)$
are also contained in the reduced sub-box ${\tilde {\cal B}}_l^\star \subseteq {{\tilde {\cal B}}_l} $, where
\begin{align}
{\bf{\tilde b}}_l^\star = {{{\bf{\tilde c}}}_l} - \sum\nolimits_{k = 1}^K {\bar \mu _k} \left( {{{\tilde c}_{l,k}} - {{\tilde b}_{l,k}}} \right){{\bf{e}}_k},\label{eq41}\\
{\bf{\tilde c}}_l^\star = {\bf{\tilde b}}_l^\star + \sum\nolimits_{k = 1}^K {\bar v _k} \left( {{{\tilde c}_{l,k}} - \tilde b_{l,k}^\star} \right){{\bf{e}}_k},\label{eq42}
\end{align}
with ${\tilde c}_{l,k}$, ${\tilde b}_{l,k}$, and  ${\tilde b}_{l,k}^\star$ being the $k$-th elements of  ${{{\bf{\tilde c}}}_l}$, ${{{\bf{\tilde b}}}_l}$, ${{{\bf{\tilde b}}}_l^\star}$, respectively, and $\bar \mu  _k $ and $\bar v  _k $ are given by
\begin{align}
\!\!{{\bar \mu }_k} &\!=\! \max\! \left\{ {\mu \left| \begin{array}{l}
0 \le \mu  \le 1,\\
{\cal L}\left( {{{{\bf{\tilde c}}}_l} - \mu \left( {{{\tilde c}_{l,k}} - {{\tilde b}_{l,k}}} \right){{\bf{e}}_k}} \right)\! \ge \!f_{\cal N}^{\min }
\end{array} \right.} \right\},\nonumber\\
\!\!{{\bar v}_k}&\! =\! \max\! \left\{ {v\left| \begin{array}{l}
0 \le v \le 1,\\
{\cal L}\left( {{\bf{\tilde b}}_l^ \star \! +\! v\left( {{{\tilde c}_{l,k}} - \tilde b_{l,k}^ \star } \right){{\bf{e}}_k}} \right) \!\le\! {f^{{\rm{ub}}}}\left( {{{\widetilde {\cal B}}_l}} \right)
\end{array} \right.} \!\!\!\!\!\right\}.
\end{align}
Note that the closed-form expressions of $\bar \mu  _k $ and $\bar v  _k $ can be obtained in special cases,~e.g., for the weighted sum function with ${\cal L}\left( {\bf{z}} \right) = \sum\nolimits_{k = 1}^K {{\tilde\alpha _k}{z_{lk}}} $  (with ${\tilde\alpha _k}>0$) for $l=1, 2$, we have
\begin{align}
\bar \mu  _k&  = \min \left\{ {1,\frac{{\sum\nolimits_{k = 1}^K {{\tilde\alpha _k}{{\tilde c}_{l,k}}}  - f_{\cal N}^{\min }}}{{{\tilde\alpha _k}\left( {{{\tilde c}_{l,k}} - {{\tilde b}_{l,k}}} \right)}}} \right\}, \\
\bar v _k & = \min \left\{ {1,\frac{{{f^{{\rm{ub}}}}\left( {\tilde{\cal  B}_l}  \right) - \sum\nolimits_{k = 1}^K {{\tilde\alpha _k}{{\tilde b}_{l,k}^\star}} }}{{{\tilde\alpha _k}\left( {{{\tilde c}_{l,k}} - {{\tilde b}_{l,k}}} \right)}}} \right\}.
\end{align}

\subsubsection{\bf Bounding}
At the end of this iteration, we search for a feasible solution in one of the new boxes ${\tilde {\cal B}}_l^\star$ and use it for improving $f^{\rm min}_{\cal N}$ and reducing $f^{\rm max}_{\cal N}$.
A popular scheme for achieving this is to project the upper vertex of ${\tilde {\cal B}}_l^\star$ onto the boundary \cite{bjornson2012robust}. However, this can potentially increase complexity, since the projection step usually requires a bisection method and each search entails conducting a feasibility check.
To reduce the complexity, with the updated box set, ${\cal N} = \left( {{\cal N}\backslash {{\cal B}_{\max }}} \right) \cup \left\{ {{\tilde{\cal  B}_1^\star},
{\tilde{\cal  B}_2^\star}} \right\}$, we update $f^{\rm min}_{\cal N}$ and $f^{\rm max}_{\cal N}$ by:
\begin{align}
f_{\cal N}^{{\rm{min}}}& = \max \left\{ {f_{\cal N}^{{\rm{min}}},{\cal L}\left( {{\bf{\tilde b}}_2^\star} \right)} \right\}, \label{eq46}\\
f_{\cal N}^{{\rm{max}}} &= \mathop {\arg \max }\limits_{{\cal B} \in {\cal N}} {f^{{\rm{ub}}}}\left( {\cal B} \right).\label{eq47}
\end{align}

\begin{remark} Note that in the reduction step, it is obvious that ${\bf \tilde b}_1^\star$ belongs to the feasible region. However,  we must check whether  ${\bf \tilde b}_2^\star$ belongs to the feasible region. In other words, we need to check whether we can find the solutions of the remaining optimization variables in problems {\bf P2a-MO} or {\bf P2b-MO} to make them feasible when ${\bf z}={\bf \tilde b}_2^\star$.
Specifically, when  using the BRB to solve problem {\bf P2a-MO}, the corresponding feasibility check problem is given by
\begin{align}
{\rm Find}&\; {\bf p}\label{eqp2aF1}\tag{\bf P2a-F}\\
{\rm s.t.}\;&\frac{{{2^{z_{qib}^{{\rm N},q}}} - 1}}{{{\gamma _{qqb}}}} \le   {p_{qb}},\nonumber\\
&\frac{{\left( {{2^{z_{qib}^{{\rm N},i}}} - 1} \right)\left( {{\gamma _{iqb}}{p_{qb}} + 1} \right)}}{{{\gamma _{iib}}}} \le {p_{ib}},\forall b \in {{\cal R}_r}, \eqref{eq31}.\nonumber
\end{align}
As for problem {\bf P2b-MO},  the corresponding feasibility check problem is given by
\begin{align}
 {\rm{Find}}&\;{\bf{p}}\label{eqp2aF2}\tag{\bf P2b-F}\\
{\rm{s}}.{\rm{t}}.\;& \frac{{\left( {{2^{z_{qb}^{\rm{S}}}} - 1} \right)\left( {\sum\nolimits_{i \ne q}^Q {{\gamma _{qib}}{p_{ib}}}  + 1} \right)}}{{{\gamma _{qqb}}}} \le {p_{qb}},\forall b \in {{\cal R}_r}, \eqref{eq32}.\nonumber
\end{align}
\end{remark}
Since the constraints are affine functions and ${z_{qib}^{{\rm N},q}}$, ${z_{qib}^{{\rm N},i}}$, and ${z_{qb}^{\rm{S}}}$ are constant values determined by the lower vertex of the corresponding box,  both \ref{eqp2aF1} and \ref{eqp2aF2} are convex optimization problems. Then, they can be efficiently solved by using the Matlab toolbox CVX \cite{grant2015cvx1}.
\hfill $\square$

Finally, the details of applying BRB to solve the MO problem are summarized in Algorithm 1.
Note that although this algorithm can obtain the $\varepsilon$-accuracy globally optimal solution at a lower complexity than the exhaustive search, the DPMO algorithm still imposes exponentially increasing complexity and cannot be applied to the general case involving rate constraints.
\begin{algorithm}[t]
\caption{Optimal solutions to {\bf P2a-MO} (or {\bf P2b-MO})}\label{algorithm2}
 {{
\begin{algorithmic}[1]
 \STATE Initialize ${\cal B}_0=\left[{\bf 0},{\bf z}_0\right]$, where the elements of ${\bf z}_0$ are set as
$z_{qib}^{{\rm N},q} ={\log _2}\left( {1 + {\gamma _{qqb}}{p_r}} \right)$, $z_{qib}^{{\rm N},i}={\log _2}\left( {1 + {\gamma _{iib}}{p_r}} \right)$ for problem \ref{p2bMO}
(or $z_{qb}^{{\rm S}} ={\log _2}\left( {1 + {\gamma _{qqb}}{p_r}} \right)$ for problem \ref{p2bMO})
\STATE Set ${\cal N}=\left\{{\cal B}_0\right\}$, and $f^{\rm min}_{\cal N} = {\cal L}^{\rm X}\left( {\bf{0}} \right) = 0$  and $f^{\rm max}_{\cal N}= {\cal L}^{\rm X}\left( {\bf{z }}_0 \right)$ where ``${\rm X}={\rm N}$''  for \ref{p2aMO} and ``${\rm X}={\rm S}$'' for \ref{p2bMO}
   \WHILE {$f^{\rm max}_{\cal N}>(1+\varepsilon)f^{\rm min}_{\cal N}$ and below the maximum number of iterations}
    \STATE Find ${{\cal B}_{\max }} $ by \eqref{eq31a}
    \FOR{$l=1,2$}
    \STATE Create new box ${{\tilde {\cal B}}_l}$ by \eqref{eq39}; Set ${f^{{\rm{ub}}}}\left( {{{\tilde {\cal B}}_l}} \right)$ by   \eqref{eq40}
    \STATE {\bf if} {$f^{\rm min}_{\cal N} \le {f^{{\rm{ub}}}}\left( {{{\tilde {\cal B}}_l}} \right)$}: \\
     $\quad$ Calculate  ${{\tilde {\cal B}}_l^\star}$ by utilizing  \eqref{eq41} and \eqref{eq42} for ${{\tilde {\cal B}}_l}$;
      \\{\bf else:} set ${{\tilde {\cal B}}_l^\star}=\emptyset$
      {\bf end if}
    \ENDFOR
    \STATE Check feasibility of ${\bf{\tilde b}}_2^\star$ by solving  {\bf P2a-F}  for  {\bf P2a-MO} (or by solving  {\bf P2b-F}  for  {\bf P2b-MO})
    \STATE {\bf if} feasible:  \\
    $\quad$ Update the lower bound $f^{\rm min}_{\cal N}$ by \eqref{eq46};
    \\{\bf else}:  set ${{\tilde {\cal B}}_2^\star}=\emptyset$
     {\bf end if}
    \STATE Update box set ${\cal N} = \left( {{\cal N}\backslash {{\cal B}_{\max }}} \right) \cup \left\{ {{\tilde{\cal  B}_1^\star},
{\tilde{\cal  B}_2^\star}} \right\}$
 \STATE Update the  upper bound $f^{\rm max}_{\cal N}$ by \eqref{eq47}
     \ENDWHILE
\STATE  {\bf Return}  a feasible solution that can achieve $f^{\rm min}_{\cal N}$.
\end{algorithmic}
}}
\end{algorithm}

\section{Suboptimal Solution to The General Case: With Rate Constraints}

The DPMO method solves problem \ref{eqP1} in the special scenario of no rate constraints \eqref{eqP1a0}, and it exhibits exponential complexity.
To efficiently solve it for the general case with rate constraints, we employ the  SCA-SA algorithm to find the sub-optimal solution within polynomial time complexity.

\subsection{Simulated Annealing (SA) Algorithm Development}
In the face of individual rate constraints, the original problem \ref{eqP1} is hard to solve and cannot be directly divided into independent problems using the DP framework.
Therefore, we need to develop efficient sub-optimal algorithms.

To begin with, we need to decouple the optimization of the binary variables ${\bf a}^{\rm S}$ and ${\bf a}^{\rm N}$.
To do so, we apply the SA algorithm \cite{di2016sub} to find the sub-optimal solution of problem \ref{eqP1}  by randomly setting ${\bf a}^{\rm S}$ iteratively.
Specifically, we denote the randomly selected solution of  ${\bf a}^{\rm S}$ and  the corresponding objective function value in the $t$-th iteration of SA by ${\bf a}^{\rm S}_t$  and $\tilde C \left( {{\bf{a}}_t^{\rm{S}}} \right)$, respectively. Besides, $\tilde C \left( {{\bf{a}}_t^{\rm{S}}} \right)$ is obtained by solving  problem \ref{eqP1} with ${{\bf{a}}^{\rm{S}}} = {\bf{a}}_t^{\rm{S}}$, i.e.,
\begin{align}
\label{eqP1r}{{\tilde C} }\left( {{\bf{a}}_t^{\rm{S}}} \right)\buildrel \Delta \over = \mathop {\max }\limits_{{{\bf{a}}^{\rm{S}}} = {\bf{a}}_t^{\rm{S}},{{\bf{a}}^{\rm{N}}},{\bf{p}}} &\;\sum\limits_{r = 1}^R {{C_r}}  \tag{{\bf{P3}}}\\
{\rm{s}}.{\rm{t}}.\quad&\eqref{eqP1a0}-\eqref{eqP1c}.\nonumber
\end{align}
The suboptimal solution to \ref{eqP1r} can be obtained by applying the SCA algorithm, which will be discussed in Section \ref{eqsubSCA}.

After solving problem \ref{eqP1r}, the current optimum user accommodation and MA scheme of ${\bf a}^{\rm S}$ can be replaced as follows:
\begin{align}
{\bf{a}}_{ \star }^{\rm S} \leftarrow \left\{ {\begin{array}{*{20}{l}}
{{\bf{a}}_t^{\rm{S}},\;{\rm{if}}\;{{\bar C}^ \star } \le {{\tilde C} }\left( {{\bf{a}}_t^{\rm{S}}} \right),}\\
{{\bf{a}}_{ \star }^{\rm S},{\rm{otherwise}}},
\end{array}} \right. \label{astar50}
\end{align}
where  ${\bf a}_\star^{\rm S}$ is the current optimal solution of ${\bf a}^{\rm S}$ that can achieve the maximum objective value
${{\bar C}^ \star }$ in the previous iterations of SA, i.e., ${\bar C^ \star } = \mathop {\max }\nolimits_{0 \le i \le t - 1} \tilde C\left( {{\bf{a}}_i^{\rm{S}}} \right)$.

Then, we iteratively explore other solutions of ${\bf a}^{\rm S}$ to potentially improve the value of the objective function. To do so, we denote an integer index $x$, which is randomly uniformly generated from the set $\left\{ {1,2,\cdots, R} \right\}$. Consequently, we update  ${\bf{a}}^{\rm{S}}$ in the $(t+1)$-th iteration  as follows:
\begin{align}
{\left[ {{\bf{a}}_{t + 1}^{\rm{S}}} \right]_r} = \left\{ {\begin{array}{*{20}{l}}
{{{\left[ {{\bf{\tilde a}}_t^{\rm{S}}} \right]}_{r}},\;{\rm{if}}\;r \ne x,}\\
{1 - {{\left[ {{\bf{\tilde a}}_t^{\rm{S}}} \right]}_{r}},{\rm{if}}\;r = x,}
\end{array}} \right.\label{astar51}
\end{align}
for $1\le r\le R$, where ${{{\left[ {{\bf{a}}_{t+1}^{\rm{S}}} \right]}_{r}}}$ and ${{{\left[ {{\bf{\tilde a}}_t^{\rm{S}}} \right]}_{r}}}$ are the $r$-th elements of vectors ${\bf{ a}}_{t+1}^{\rm{S}}$ and  ${\bf{\tilde a}}_{t}^{\rm{S}}$, respectively.
Besides, ${\bf{\tilde a}}_t^{\rm{S}}$ is the  solution updated by comparing  ${{\bar C}^ \star }$ and ${{\tilde C} }\left( {{\bf{a}}_t^{\rm{S}}} \right)$, i.e.,
\begin{align}
{\bf{\tilde a}}_t^{\rm{S}} = \left\{ {\begin{array}{*{20}{l}}
{{\bf{a}}_t^{\rm{S}},\;{\rm{if}}\;{{\bar C}^ \star } \le \tilde C\left( {{\bf{a}}_t^{\rm{S}}} \right),}\\
{{\bf{\hat a}}_t^{\rm{S}},\;{\rm{if}}\;{{\bar C}^ \star }> \tilde C\left( {{\bf{a}}_t^{\rm{S}}} \right),}
\end{array}} \right.
\end{align}
where ${\bf{\hat a}}_t^{\rm{S}} = \left\{ {\begin{array}{*{20}{l}}
{{\bf{a}}_t^{\rm{S}},{\rm{with}}\;{\rm{probability}}\;{\cal P}_t }\\
{{\bf{\tilde a}}_\star^{\rm{S}}{\rm{,{\rm{with}}\;{\rm{probability}}\; (1-{\cal P}_t)}}}
\end{array}} \right.$. Here, we define the probability by ${\cal P}_t=\exp \left( {\frac{1}{{T_e^t}}\left( {{{\tilde C} }\left( {{\bf{a}}_t^{\rm{S}}} \right) - {{\bar C}^ \star }} \right)} \right)$, where $T_e^t$ is the temperature in $t$-th iteration of the SA algorithm, which is updated by $T_e^{t+1}={\bar \zeta }T_e^t$.
The parameter $0<{\bar \zeta }<1$ is the cooling factor, which determines how the temperature gradually decreases throughout the iterations, influencing the exploration and convergence speed in the search space.
In this way, it allows the selected suboptimal ${\bf a}^{\rm S}$ a chance to enable the SA algorithm to escape poor local optima and converge to better solutions.
Note that to reduce the computational complexity, if the   randomly generated index $x$ results in
${\bf{a}}_{t + 1}^{\rm{S}}$  that has already appeared, we have to regenerate this index so that the resultant  ${\bf{a}}_{t + 1}^{\rm{S}} $ is unprecedented.

It is readily apparent that the utility will not decrease during each iteration.
Furthermore, the convergence can be guaranteed because the probability ${\cal P}_t$ decreases as the temperature decreases.
Finally, the details are summarized in Algorithm~\ref{algorithm31A}.

\subsection{Sub-Optimal Solution of Problem  \ref{eqP1r} }\label{eqsubSCA}

During each iteration of SA, we have to solve problem \ref{eqP1r} with ${{\bf{a}}^{\rm{S}}} = {\bf{a}}_t^{\rm{S}}$. However, this problem is hard to solve due to the remaining binary variables ${\bf a}^{\rm N}$.
In the following, we transform problem \ref{eqP1r} into a solvable form by introducing additional auxiliary and some necessary approximations and then apply the SCA algorithm to solve it efficiently.

Firstly,  we introduce the auxiliary optimization variables $\bf \tilde p$, whose elements include
$\left\{ {\tilde p_{qib}^{{\rm{N}},q},\tilde p_{qib}^{{\rm{N}},i}\left| {\forall b,\forall \left( {q,i} \right) \in{\cal Q}} \right.} \right\}$ and
$\left\{ {\tilde p_{qb}^{\rm{S}}\left| {\forall b,\forall q} \right.} \right\}$.
Here,  we define
$\tilde p_{qib}^{{\rm{N}},q} = a_{rqi}^{\rm{N}}{p_{qb}}$ and $\tilde p_{qib}^{{\rm{N}},i} = a_{rqi}^{\rm{N}}{p_{ib}}$ if $b \in {\cal R}_r$,
and $\tilde p_{qb}^{\rm{S}} = {{{\left[ {{\bf{ a}}_t^{\rm{S}}} \right]}_{r}}}{p_{qb}}$  if $b \in {\cal R}_r$.
Since some of the multiplicative terms consist of binary variables $ a_{rqi}^{\rm{N}}$, problem  \ref{eqP1r} is challenging to solve.
We apply the big-M method to decompose them \cite{chen2019optimal}, yielding:
\begin{align}
&{p_{qb}} - \left( {1 - a_{rqi}^{\rm{N}}} \right)P \le \tilde p_{qib}^{{\rm{N}},q} \le a_{rqi}^{\rm{N}}P,\;0 \le \tilde p_{qib}^{{\rm{N}},q} \le {p_{qb}},\nonumber\\
&\qquad\qquad\qquad\qquad\qquad{\rm{for}}\;\forall b \in {{\cal R}_r},\forall r,\forall \left( {q,i} \right) \in {\cal Q}, \label{eqintegerr1} \\
&{p_{ib}} - \left( {1 - a_{rqi}^{\rm{N}}} \right)P \le \tilde p_{qib}^{{\rm{N}},i} \le a_{rqi}^{\rm{N}}P, \;0 \le \tilde p_{qib}^{{\rm{N}},i} \le {p_{ib}}, \nonumber\\
&\qquad\qquad\qquad\qquad\qquad{\rm{for}}\;\forall b \in {{\cal R}_r},\forall r,\forall \left( {q,i} \right) \in {\cal Q}. \label{eqintegerr2}
\end{align}

Based on these auxiliary variables and the average rate defined in \eqref{eqRNOMA002} and \eqref{eqRsdma}, we have
\begin{align}
\tilde C_{rqi}^{{\rm{N}},q} &\buildrel \Delta \over = a_{rqi}^{\rm{N}}C_{rqi}^{{\rm{N}},q} = \frac{1}{{MN}}\sum\limits_{b \in {{\cal R}_r}} {{{\log }_2}\left( {1 + {\gamma _{qqb}}\tilde p_{qib}^{{\rm{N}},q}} \right)} ,\\
\tilde C_{rqi}^{{\rm{N}},i} &\buildrel \Delta \over = a_{rqi}^{\rm{N}}C_{rqi}^{{\rm{N}},i} = \frac{1}{{MN}}\sum\limits_{b \in {{\cal R}_r}} {{{\log }_2}\left( {1 + \nu _{qib}^{{\rm{N}}}} \right)}, \\
\tilde C_{rq}^{\rm{S}} &\buildrel \Delta \over = a_{t,r}^{\rm{S}}C_{rq}^{\rm{S}} = \frac{1}{{MN}}\sum\limits_{b \in {{\cal R}_r}} {{{\log }_2}\left( {1 + \nu _{qb}^{\rm{S}}} \right)},
\end{align}
where $
\nu _{qib}^{{\rm{N}}} = \frac{{{\gamma _{iib}}\tilde p_{qib}^{{\rm{N}},i}}}{{ {\gamma _{iqb}}\tilde p_{qib}^{{\rm{N}},q}}+1}$  for $b\in{\cal R}_r,\forall  r,\forall\left( {q,i} \right) \in {\cal Q}$ and $\nu _{qb}^{\rm{S}} = \frac{{{\gamma _{qqb}}\tilde p_{qb}^{\rm{S}}}}{{\sum\nolimits_{i \ne q} {{\gamma _{qib}}\tilde p_{ib}^{\rm{S}}}  + 1}}$
 for $b\in{\cal R}_r,\forall  r,\forall q$. Besides, we define $a_{t,r}^{\rm{S}}={{{\left[ {{\bf{  a}}_t^{\rm{S}}} \right]}_{r}}}$ for notation simplicity.

\begin{algorithm}[t]
\caption{SCA-SA algorithm to solve \ref{eqP1}  }\label{algorithm31A}
 {{
\begin{algorithmic}[1]
 \STATE   Initialize ${\bf{a}}_{\star }^{\rm S}={\bf 1}_{R}$; initialize ${{\bar C}^ \star } = \tilde C \left( {\bf{a}}_{\star }^{\rm S} \right) $
 by using SCA to solve problem \ref{eqP1r}; initialize $t=0$, ${\bf{a}}_{t }^{\rm S}={\bf 0}_{R}$, ${\bar \zeta }=0.92$, $T_e^t= \max \left\{ {{{\bar C}^ \star },10} \right\}$
\REPEAT
 \STATE $t\leftarrow t+1$ and calculate ${{\tilde C} }\left( {{\bf{a}}_t^{\rm{S}}} \right)$  by using SCA algorithm to solve \ref{eqP1r} (i.e., solve \ref{eqP32r} iteratively)
 \STATE Update ${\bf{a}}_{ \star }^{\rm S}$ by \eqref{astar50};   Update ${\bf{a}}_{t+1}^{\rm{S}}$ by \eqref{astar51}; Update ${{\bar C}^ \star } = \max \left\{ {{{\bar C}^ \star },\tilde C\left( {{\bf{a}}_t^{\rm{S}}} \right)} \right\}$ and $T_e^{t+1}={\bar \zeta }T_e^t$
\UNTIL Convergence
\STATE  {\bf Return} the solutions of ${{\bf{a}}^{\rm{S}}}$, ${{\bf{a}}^{\rm{N}}}$, and ${\bf{p}}$.
\end{algorithmic}
}}
\end{algorithm}

By substituting them into problem \ref{eqP1r},  the objective function and the achievable transmission rate of the $q$-th user defined in the individual rate constraint \eqref{eqP1a0} can be reformulated as:
 \begin{align}
\tilde C\left( {{\bf{\tilde p}},{\bm{\nu }}} \right)&= \sum\limits_{r = 1}^R {{C_r}} \nonumber\\
&= \sum\limits_{q = 1}^{Q-1} {\sum\limits_{i = q + 1}^Q {\left( {{\alpha _q}\tilde C_{rqi}^{{\rm{N}},q} + {\alpha _i}\tilde C_{rqi}^{{\rm{N}},i}} \right)} }  + \sum\limits_{q = 1}^Q {{\alpha _q}\tilde C_{rq}^{\rm{S}}}, \\
{{\tilde C}_q}\left( {{\bf{\tilde p}},{\bm{\nu }}} \right)&= \sum\limits_{r = 1}^R {\left( {\sum\limits_{i = 1}^{q - 1} {\tilde C_{riq}^{{\rm{N}},q}}  + \sum\limits_{i = q + 1}^Q {\tilde C_{rqi}^{{\rm{N}},q} + } \tilde C_{rq}^{\rm{S}}} \right)},
\end{align}
where $\bm \nu$ is a vector consisting of these elements $\left\{ {\nu _{qib}^{\rm{N}}\left| {\forall b \in {{\cal R}_r},\forall r,\forall \left( {q,i} \right) \in {\cal Q}} \right.} \right\}$ and $\left\{ {\nu _{qb}^{\rm{S}}\left| {\forall b \in {{\cal R}_r},\forall r,\forall q} \right.} \right\}$.
Due to the existence of the individual QoS constraint, we introduce the following constraints to guarantee that each user can gain channel access at least once for information transmission if ${C_{\min,q}}>0$, i.e.,
\begin{align}
\sum\nolimits_{r = 1}^R {\left( {\sum\nolimits_{i = 1}^{q - 1} {a_{riq}^{\rm{N}} + \sum\nolimits_{i = q + 1}^Q {a_{rqi}^{\rm{N}} + a_{t,r}^{\rm{S}}} } } \right)} \ge {{\rm{I}}_q},\label{eqc61}
\end{align}
where ${{\rm{I}}_q}=\left\{0,1\right\}$ is a constant dependent on the minimal QoS requirement, i.e., ${{\rm{I}}_q}=1$ if ${C_{\min,q}}>0$, otherwise ${{\rm{I}}_q}=0$.

To address the binary integer constraint \eqref{eqP1c} in problem \ref{eqP1r}, we introduce the following joint continuous equivalent constraints to replace it:
\begin{align}
&0 \le a_{rqi}^{\rm{N}} \le 1,\forall r,\forall \left( {q,i} \right) \in {\cal Q},\label{eqAapproximate1}\\
&{\sum\nolimits_{r = 1}^R {\sum\nolimits_{q = 1}^{Q-1} {\sum\nolimits_{i = q + 1}^Q {\left( {a_{rqi}^{\rm{N}} - {{\left( {a_{rqi}^{\rm{N}}} \right)}^2}} \right)} } } }\le0.
\label{eqAapproximate2}
\end{align}
With constraints \eqref{eqAapproximate1} and \eqref{eqAapproximate2}, the binary variables $a_{rqi}^{\rm{N}}$ can be converted into continuous optimization variables. Although there are only continuous constraints and no direct integer constraints \eqref{eqP1c}, the joint constraints (54) and (55) effectively restrict the solution to integer values. Thus, this step is an equivalent substitution.

However, constraint \eqref{eqAapproximate2} is non-convex, which still makes  \ref{eqP1r} hard to solve.
Therefore, we introduce a penalty factor $\eta$ to relax \eqref{eqAapproximate2} in the objective function.
By further combining it with the above additional constraints and variables, we can reformulate problem   \ref{eqP1r} into the following form:
\begin{subequations}
\begin{align} \label{eqP3r} \mathop {\max }\limits_{{{\bf{a}}^{\rm{N}}},{\bf{p}},{\bf{\tilde p}},{\bm{\nu }}}&\tilde  C\left({\bf{\tilde p}},{\bm{\nu }} \right) + \eta \left( { \tilde G\left( {{{\bf{a}}^{\rm N}}}  \right)-G\left( {{{\bf{a}}^{\rm N}}} \right)} \right)\tag{{\bf{P3-1}}}\\
{\rm{s}}{\rm{.t}}{\rm{.}}\;&\tilde C_q{\left( {\bf{\tilde p}},{\bm{\nu }} \right)} \ge {C_{\min ,q}}, \forall q,\label{eq56a}\\
&V_{qib}^{\rm{N}}\left( {{\bf{\tilde p}},{\bm{\nu }}} \right) - \tilde V_{qib}^{\rm{N}}\left( {{\bf{\tilde p}},{\bm{\nu }}} \right) \le {\gamma _{iib}}\tilde p_{qib}^{{\rm{N}},i},\nonumber\\
&\qquad\qquad\qquad\qquad \forall b \in {{\cal R}_r},\forall r,\forall \left( {q,i} \right) \in {\cal Q},\\
&{\gamma _{qqb}}\tilde p_{rqi}^{{\rm{N,}}q} - {\gamma _{qib}}\tilde p_{rqi}^{{\rm{N,}}i} \le 0,\nonumber\\
&\qquad\qquad\qquad\qquad \forall b \in {{\cal R}_r},\forall r,\forall \left( {q,i} \right) \in {\cal Q},\label{eq56d}\\
&V_{qb}^{\rm{S}}\left( {{\bf{\tilde p}},{\bm{\nu }}} \right) - \tilde V_{qb}^{\rm{S}}\left( {{\bf{\tilde p}},{\bm{\nu }}} \right) \le {\gamma _{qqb}}\tilde p_{qb}^{\rm{S}},\nonumber\\
&\qquad\qquad\qquad\qquad\qquad\quad\;\;\; \forall b \in {{\cal R}_r},\forall r,\forall q,\\
&\eqref{eqP1a},\eqref{eqP1b0}, \eqref{eqP1d}, \eqref{eqintegerr1}, \eqref{eqintegerr2}, \eqref{eqc61}, \eqref{eqAapproximate1}, \nonumber
\end{align}
\end{subequations}
where the factor $\eta$, which should be much larger than~1~\cite{chen2019optimal,shan2024resource}, is used to penalize the objective function in cases where $a_{rqi}^{\rm N}$ is neither 0 nor 1;   \eqref{eq56d} is due to \eqref{eqP1a1}; and
\begin{align}
&G\left( {{{\bf{a}}^{\rm N}}} \right)= \sum\nolimits_{r = 1}^R {\sum\nolimits_{q = 1}^{Q-1}{\sum\nolimits_{i = q + 1}^Q {a_{rqi}^{\rm{N}}} } },\\
 &\tilde G\left( {{{\bf{a}}^{\rm N}}} \right) = \sum\nolimits_{r = 1}^R {\sum\nolimits_{q = 1}^{Q-1} {\sum\nolimits_{i = q + 1}^Q {{{\left( {a_{rqi}^{\rm{N}}} \right)}^2}} } }, \\
&V_{qib}^{\rm{N}}\left( {{\bf{\tilde p}},{\bm{\nu }}} \right) =\frac{1}{4} {\left( {\nu _{qib}^{\rm{N}} + {\gamma _{iqb}}\tilde p_{qib}^{{\rm{N}},q} + 1} \right)^2},\\
&\tilde V_{qib}^{\rm{N}}\left( {{\bf{\tilde p}},{\bm{\nu }}} \right) = \frac{1}{4}{\left( {\nu _{qib}^{\rm{N}} - {\gamma _{iqb}}\tilde p_{qib}^{{\rm{N}},q} - 1} \right)^2},\\
&V_{qb}^{\rm{S}}\left( {{\bf{\tilde p}},{\bm{\nu }}} \right) = \frac{1}{4}{\left( {\nu _{qb}^{\rm{S}} + \sum\nolimits_{i \ne q} {{\gamma _{qib}}\tilde p_{ib}^{\rm{S}}}  + 1} \right)^2},\\
&\tilde V_{qb}^{\rm{S}}\left( {{\bf{\tilde p}},{\bm{\nu }}} \right) = \frac{1}{4}{\left( {\nu _{qb}^{\rm{S}} - \sum\nolimits_{i \ne q} {{\gamma _{qib}}\tilde p_{ib}^{\rm{S}}}  - 1} \right)^2}.
 \end{align}
It is obvious that $\tilde C\left( {{\bf{\tilde p}}, {\bm{\nu }}} \right)$ and ${{\tilde C}_q}\left( {{\bf{\tilde p}},{\bm{\nu }}} \right)$ are concave functions, and $G\left( {{{\bf{a}}^{\rm N}}} \right)$, $\tilde G\left( {{{\bf{a}}^{\rm N}}} \right)$, $V_{qib}^{{\rm{N}}}\left( {{\bf{\tilde p}},{\bm{\nu }}} \right)$, $\tilde V_{qib}^{{\rm{N}}}\left( {{\bf{\tilde p}},{\bm{\nu }}} \right)$, $V_{qb}^{\rm{S}}\left( {{\bf{\tilde p}},{\bm{\nu }}} \right)$, and $\tilde V_{qb}^{\rm{S}}\left( {{\bf{\tilde p}},{\bm{\nu }}} \right)$ are affine or convex functions. Thus, problem \ref{eqP3r} belongs to the difference of convex (DC) family \cite{chen2018resource}, where the objective function or constraints can be expressed in terms of the differences between two concave/convex functions.
Hence, we can apply the SCA method to solve it iteratively. In each iteration, the Taylor series expansion is used to approximate the convex/concave terms by linear functions, thus transforming them into convex problems.
Specifically, defining the solutions of ${{\bf{a}}^{\rm{N}}}$, ${\bf{\tilde p}}$, ${\bm{\nu }}$
 in the $(\imath-1)$-th iteration by ${\bf{a}}^{\rm{N}}_\imath$, ${\bf{\tilde p}}_\imath$, ${\bm{\nu }}_\imath$, we have
\begin{align}
\tilde G\left( {{{\bf{a}}^{\rm{N}}}} \right) &\mathop  \ge \limits_{\left( \rm a \right)}  \tilde G\left( {{\bf{a}}_\imath ^{\rm{N}}} \right) + \nabla \tilde G{\left( {{\bf{a}}_\imath ^{\rm{N}}} \right)^T}\left( {{{\bf{a}}^{\rm{N}}} - {\bf{a}}_\imath ^{\rm{N}}} \right) \nonumber\\
& \buildrel \Delta \over = {{\tilde G}^{{\rm{lb}}}}\left( {{{\bf{a}}^{\rm{N}}};{\bf{a}}_\imath ^{\rm{N}}} \right),\label{eq7a71}\\
\tilde V_{qib}^{{\rm{N}}}\left( {{\bf{\tilde p}},{\bm{\nu }}} \right){\rm{ }}&\mathop  \ge \limits_{\left( {\rm{b}} \right)} \tilde V_{qib}^{{\rm{N}}}\left( {{\bf{\tilde p}}_\imath,{\bm{\nu }}_\imath} \right) + {\left[ {\begin{array}{*{20}{c}}
{\nabla {\tilde V}_{qib}^{{\rm{N}}}\left( {{{{\bf{\tilde p}}}_\imath }} \right)}\\
{\nabla {\tilde V}_{qib}^{{\rm{N}}}\left( {{{\bm{\nu }}_\imath }} \right)}
\end{array}} \right]^T}\left[ {\begin{array}{*{20}{c}}
{{\bf{\tilde p}} - {{{\bf{\tilde p}}}_\imath }}\\
{{\bm{\nu }} - {{\bm{\nu }}_\imath }}
\end{array}} \right] \nonumber\\
& \buildrel \Delta \over = \tilde V_{qib}^{{\rm{lb}},{\rm{N}}}\left( {{\bf{\tilde p}},{\bm{\nu }};{{{\bf{\tilde p}}}_\imath },{{\bm{\nu }}_\imath }} \right),\label{eq7a72}\\
\tilde V_{qb}^{\rm{S}}\left( {{\bf{\tilde p}},{\bm{\nu }}} \right) & \mathop  \ge \limits_{\left( {\rm{c}} \right)} \tilde V_{qb}^{\rm{S}}\left( {{\bf{\tilde p}}_\imath,{\bm{\nu }}_\imath} \right) + {\left[ {\begin{array}{*{20}{c}}
{\nabla {\tilde V}_{qb}^{\rm{S}}\left( {{{{\bf{\tilde p}}}_\imath }} \right)}\\
{\nabla {\tilde V}_{qb}^{\rm{S}}\left( {{{\bm{\nu }}_\imath }} \right)}
\end{array}} \right]^T}\left[ {\begin{array}{*{20}{c}}
{{\bf{\tilde p}} - {{{\bf{\tilde p}}}_\imath }}\\
{{\bm{\nu }} - {{\bm{\nu }}_\imath }}
\end{array}} \right] \nonumber\\
&\buildrel \Delta \over = \tilde V_{qb}^{{\rm{lb}},{\rm{S}}}\left( {{\bf{\tilde p}},{\bm{\nu }};{{{\bf{\tilde p}}}_\imath },{{\bm{\nu }}_\imath }} \right),\label{eq7a73}
 \end{align}
 where
$\nabla \tilde G\left( {{\bf{a}}_\imath ^{\rm{N}}} \right) = \frac{{\partial \tilde G\left( {{{\bf{a}}^{\rm{N}}}} \right)}}{{\partial {{\bf{a}}^{\rm{N}}}}}\left| {_{{{\bf{a}}^{\rm{N}}} = {\bf{a}}_\imath ^{\rm{N}}}} \right.$,
$\nabla {\tilde V}_{qib}^{{\rm{N}}}\left( {{{{\bf{\tilde p}}}_\imath }} \right) = \frac{{\partial \tilde V_{qib}^{{\rm{N}}}\left( {{\bf{\tilde p}},{\bm{\nu }}} \right)}}{{\partial {\bf{\tilde p}}}}\left| {_{{\bf{\tilde p}} = {{{\bf{\tilde p}}}_\imath },{\bm{\nu }} = {{\bm{\nu }}_\imath }}} \right.$,
$\nabla {\tilde V}_{qib}^{{\rm{N}}}\left( {{{\bm{\nu }}_\imath }} \right) = \frac{{\partial {\tilde V}_{qib}^{{\rm{N}}}\left( {{\bf{\tilde p}},{\bm{\nu }}} \right)}}{{\partial {\bm{\nu }}}}\left| {_{{\bf{\tilde p}} = {{{\bf{\tilde p}}}_\imath },{\bm{\nu }} = {{\bm{\nu }}_\imath }}} \right.$,
$\nabla {\tilde V}_{qb}^{\rm{S}}\left( {{{{\bf{\tilde p}}}_\imath }} \right) = \frac{{\partial \tilde V_{qb}^{\rm{S}}\left( {{\bf{\tilde p}},{\bm{\nu }}} \right)}}{{\partial {\bf{\tilde p}}}}\left| {_{{\bf{\tilde p}} = {{{\bf{\tilde p}}}_\imath },{\bm{\nu }} = {{\bm{\nu }}_\imath }}} \right.$,
and
$\nabla {\tilde V}_{qb}^{\rm{S}}\left( {{{\bm{\nu }}_\imath }} \right) = \frac{{\partial \tilde V_{qb}^{\rm{S}}\left( {{\bf{\tilde p}},{\bm{\nu }}} \right)}}{{\partial {\bm{\nu }}}}\left| {_{{\bf{\tilde p}} = {{{\bf{\tilde p}}}_\imath },{\bm{\nu }} = {{\bm{\nu }}_\imath }}} \right. $ are constant gradients of the functions $
\tilde G\left( {{{\bf{a}}^{\rm{N}}}} \right)$, ${\tilde V}_{qib}^{{\rm{N}}}\left( {{\bf{\tilde p}},{\bm{\nu }}} \right)$, and ${\tilde V}_{qb}^{\rm{S}}\left( {{\bf{\tilde p}},{\bm{\nu }}} \right)$ with respect to $\bf a$, $\tilde {\bf p}$, and $\bm \nu$, respectively. Here,  the inequalities at (a) in \eqref{eq7a71}, (b) in \eqref{eq7a72}, and (c) in \eqref{eq7a73}, are due to the semidefinite Hessian matrix of the concave function, which makes the right-hand side lower than the left-hand side \cite{chen2018resource}.

Therefore, by harnessing the lower bounds defined in  \eqref{eq7a71}, \eqref{eq7a72}, and \eqref{eq7a73}, problem \ref{eqP3r} can be solved by iteratively solving the following problem
\begin{subequations}
\begin{align}
 \label{eqP32r}
\mathop {\max }\limits_{{{\bf{a}}^{\rm{N}}},{\bf{p}},{\bf{\tilde p}},{\bm{\nu }}}& C\left( {{\bf{\tilde p}},{\bm{\nu }}} \right) + \eta \left( {{{\tilde G}^{{\rm{lb}}}}\left( {{{\bf{a}}^{\rm{N}}};{\bf{a}}_\imath ^{\rm{N}}} \right) - G\left( {{{\bf{a}}^{\rm N}}} \right)} \right)\tag{{\bf{P3-2}}}\\
{\rm{s}}.{\rm{t}}. \;& V_{qib}^{\rm{N}}\left( {{\bf{\tilde p}},{\bm{\nu }}} \right) - \tilde V_{qib}^{{\rm{lb}},{\rm{N}}}\left( {{\bf{\tilde p}},{\bm\nu} ;{{{\bf{\tilde p}}}_\imath },{{\bm\nu} _\imath }} \right) \le {\gamma _{iib}}\tilde p_{qib}^{{\rm{N}},i},\nonumber\\
&\qquad\qquad\qquad\qquad \forall b \in {{\cal R}_r},\forall r,\forall \left( {q,i} \right) \in {\cal Q},\\
&V_{qb}^{\rm{S}}\left( {{\bf{\tilde p}},{\bm{\nu }}} \right) - \tilde V_{qb}^{{\rm{lb}},{\rm{S}}}\left( {{\bf{\tilde p}},{\bm{\nu }};{{{\bf{\tilde p}}}_\imath },{{\bm{\nu }}_\imath }} \right) \le {\gamma _{qqb}}\tilde p_{qb}^{\rm{S}},\nonumber\\
&\qquad\qquad\qquad\qquad\qquad\qquad \forall b \in {{\cal R}_r},\forall r,q,\\
&\eqref{eqP1a},\eqref{eqP1b0}, \eqref{eqP1d}, \eqref{eqintegerr1}, \eqref{eqintegerr2}, \eqref{eqc61}, \eqref{eqAapproximate1},\eqref{eq56a},\eqref{eq56d}. \nonumber
 \end{align}
\end{subequations}
Obviously, problem \ref{eqP32r} is convex due to the concave objective function and convex constraints.  Therefore, it can be solved optimally by using the Matlab toolbox CVX \cite{grant2015cvx1}. Then, we repeat this approximation and the convergence can be guaranteed due to the iteratively increased objective function.
In summary, problem \ref{eqP1r} is relaxed to \ref{eqP3r}, and its approximated solution can be obtained using the SCA algorithm by iteratively solving problem \ref{eqP32r}.

\subsection{Complexity Analysis}

In this part, we analyze the complexity of Algorithm~\ref{algorithm31A}, which consists of an inner loop for the SCA algorithm and an outer loop for the SA algorithm. The primary computational complexity arises from solving the convex optimization problem \ref{eqP32r} within the inner loop, because we provide a closed-form update scheme in \eqref{astar50} for the outer loop.
Specifically, from \cite{nesterov1994interior},  the complexity of solving a convex optimization problem can be expressed as ${\cal O}\left( {{\cal N}_{\rm{C}}^{1.5}{\cal N}_{\rm{D}}^{2}} \right)\ln \left( {\frac{{{\cal W}\left( {\cal P} \right)}}{{\mathord{\buildrel{\lower3pt\hbox{$\scriptscriptstyle\frown$}}
\over \epsilon } }}} \right)$, where $\cal P$ and ${\mathord{\buildrel{\lower3pt\hbox{$\scriptscriptstyle\frown$}}
\over \varepsilon } }$ represent any problem instance from the optimization family and the accuracy parameter, respectively.
Besides,  ${{\cal W}\left( {\cal P} \right)}$ is a certain data-dependent scale factor, and  the quantity $\ln \left( {\frac{{{\cal W}\left( {\cal P} \right)}}{{\mathord{\buildrel{\lower3pt\hbox{$\scriptscriptstyle\frown$}} \over \varepsilon } }}} \right)$ represents the number of accuracy digits in ${\mathord{\buildrel{\lower3pt\hbox{$\scriptscriptstyle\frown$}}
\over \varepsilon } }$-solution.
 Additionally, ${{\cal N}_{\rm C}} = \frac{{Q\left( {Q - 1} \right)}}{2}\left( {9MN + R} \right) + Q\left( {MN + 2} \right)$ and
${{\cal N}_{\rm D}} = \frac{{Q\left( {Q - 1} \right)}}{2}\left( {3MN + R} \right) + 3QMN$ represents the number of constraints and the dimension of optimization variables of problem \ref{eqP32r}.
Next, Let ${{\cal N}_{\rm{SC}}}$ and ${{\cal N}_{\rm{SA}}}$ denote the number of iterations for the SCA and SA algorithms, respectively. The overall complexity is given by
${\cal O}\left( {{{\cal N}_{\rm{SC}}}{{\cal N}_{\rm{SA}}}{\cal N}_{\rm{C}}^{1.5}{\cal N}_{\rm{D}}^2} \right)\ln \left( {\frac{{{\cal W}\left( {\cal P} \right)}}{{\mathord{\buildrel{\lower3pt\hbox{$\scriptscriptstyle\frown$}}
\over \varepsilon } }}} \right)$.
This implies that  Algorithm \ref{algorithm31A} can solve  \ref{eqP1} within polynomial complexity.

\section{Simulation Results}

In this section, we provide simulation results to validate the effectiveness of the proposed algorithms.
In the absence of specific elaboration, the noise powers $\varsigma_q$ of all users are set to $-108$ dBm and the channel $h_{q,\ell}$ obeys independent Rayleigh fading with power $\sigma_q = 10^{\frac{{\rm PL}_q}{10}}$, where $10^{\frac{{\rm PL}_q}{10}}$ is the path-loss. Besides, we set ${\rm PL}_q$  as ${\rm PL}_q=-30.5-36.7\log_{10} \frac{D_q}{1m}$ \cite{mohammadi2022cell}, where $D_q$ is the distance between user $q$ and BS, which is generated from the uniform distribution between $200$~m and $500$~m.
Then, we re-order it as a descending order, i.e., ${\sigma _q} \ge {\sigma _{\tilde q}}$ if $1\le q<\tilde q\le Q$ for simplicity.
The angle of departure is modeled by ${\theta _{q,\ell }} = {{\bar \theta }_q} + {\hat \theta }_{q,\ell}$, where ${{\bar \theta }_q}$ and ${\hat \theta }_{q,\ell}$ are randomly generated from the uniform distributions of ${\cal U}[0,\pi/2]$ and ${\cal U}[-\pi/8,\pi/8]$, respectively.
Furthermore, the Doppler $\upsilon _{q,\ell}$  and delay $\tau _{q,\ell}$ are randomly generated from the following uniform distributions ${\cal U}[-v_{\rm max},v_{\rm max}]$ and ${\cal U}[0,\tau_{\rm max}]$, where the maximum Doppler and delay of each user are set to  $\upsilon_{\rm max}=0.5\Delta_f$ and  $\tau_{\rm max}=0.5/\Delta_f$, respectively.
 Here, the subcarrier bandwidth is set to $\Delta _f=$ 15 KHz. The parameter $N_{q,\ell}$ is set to 5.
The minimal transmission rate and the corresponding weight of each user are set to $C_{{\rm min},q}=C_{\rm min} $ (bits/s/Hz)  and $\alpha_q=1$, respectively.
Besides, the total number of propagation paths of each user is set to $L_q=5$.
The accuracy parameter and penalty factor are set to $\varepsilon=1.05$  and $\eta=1000$, respectively.
Note that due to having the rate constraints, the problem formulated may either become unsolvable or the solutions may not meet these constraints.
Then, it will be deemed to be an outage case, where we regard the total rate of this channel realization as being zero.
Finally, all results are averaged over 100 Monte Carlo trials. Additionally, we compare the performances of the proposed optimal DPMO in Algorithm \ref{algorithmDPMO1} and the suboptimal SCA-SA in Algorithm \ref{algorithm31A} with the following baselines.

 \begin{figure}[t] \centering
   \centering
  \includegraphics[width=0.48\textwidth]{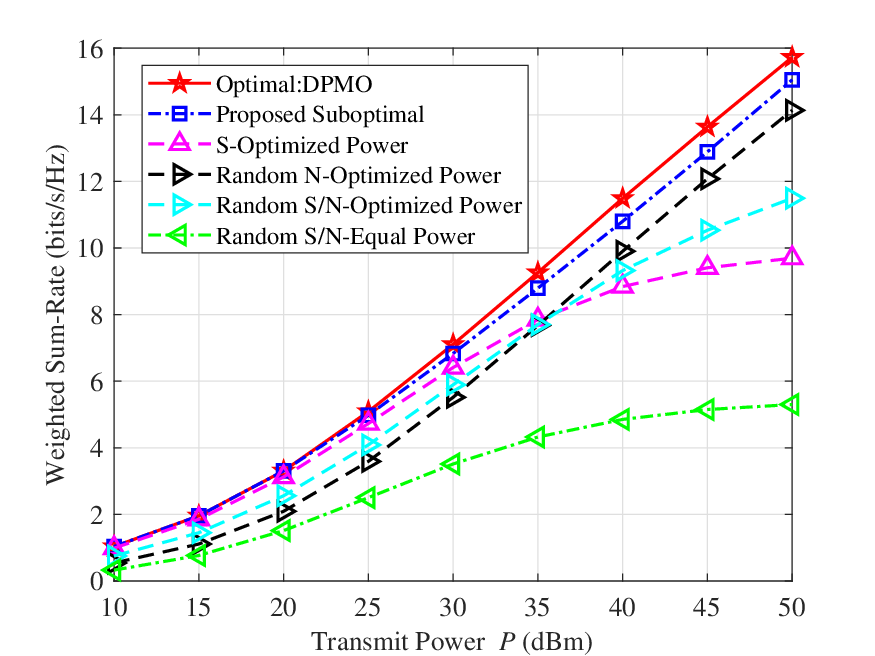}
    \caption{The weighted sum-rate of all users versus transmission power without considering QoS constraint: $Q=3$, $D=10$, $M=2$, $N=4$, $\delta_M=1$, and $\delta_N=2$.  }\label{figsim1}
 \end{figure}

 \begin{figure*}[t] \centering
 \begin{minipage}{.48 \textwidth}
   \centering
    \includegraphics[width=\textwidth,]{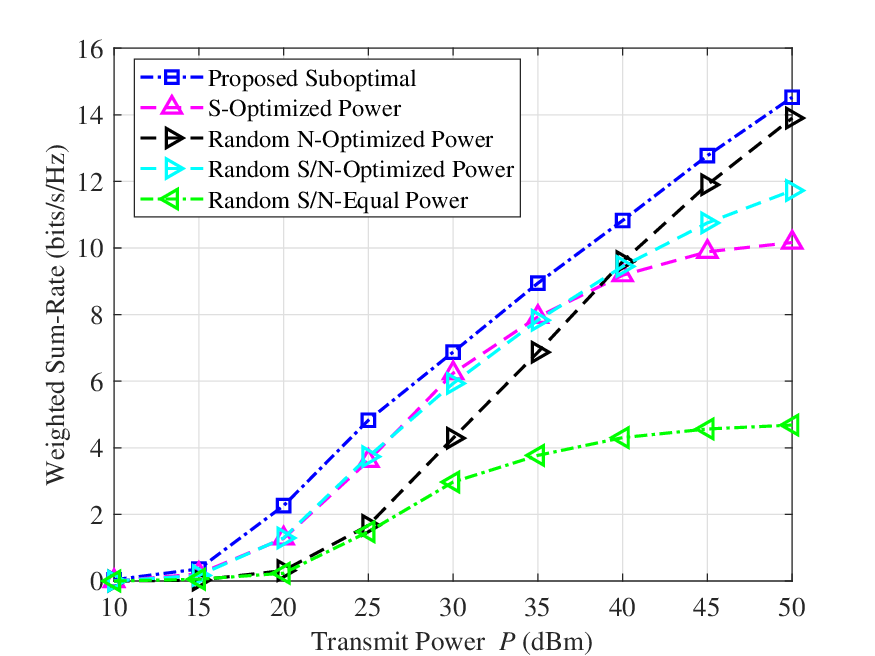}
    \subcaption{Weighted Sum-Rate}\label{figsim21}
  \end{minipage}
    \begin{minipage}{.48 \textwidth}
    \centering
    \includegraphics[width=\textwidth]{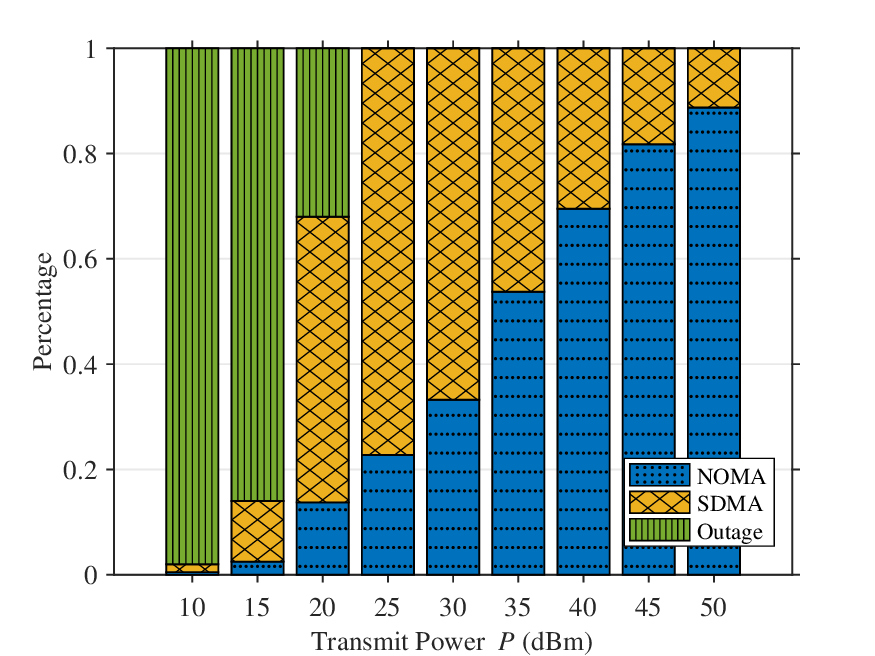}
   \subcaption{Occurrence Percentage}\label{figsim22}
   \end{minipage}
    \caption{The weighted sum-rate of all users versus transmission power with QoS constraint: $C_{\rm min}= 0.6$, $Q=3$, $D=10$, $M=2$, $N=4$, $\delta_M=1$, and $\delta_N=2$.}\label{figsim2}
 \end{figure*}

\begin{itemize}
\item {{S-Optimized Power}}: We implement SDMA for all DD RSs and then modify the SCA algorithm to solve the remaining power allocation problem.

\item {\emph{Random N-Optimized Power}}: We randomly select two users for each RS using the NOMA protocol, ensuring they satisfy constraint \eqref{eqc61}. Then, we modify the SCA algorithm to solve the power allocation problem.
\item {{Random S/N-Optimized Power}}: We randomly choose between SDMA or NOMA for each DD RS, then select users for each RS while meeting constraint \eqref{eqc61}. Next, we modify the SCA algorithm to address the power allocation problem.

\item{{Random S/N-Equal Power}}: We randomly choose between SDMA or NOMA for each DD RS, then select users for each RS while meeting constraint \eqref{eqc61}. Next, we apply equal power allocation to the users in each RS
\end{itemize}

Fig. \ref{figsim1} investigates the impact of transmission power on the weighted sum-rate without considering the individual rate constraint.
Firstly, we observe that the performances of all schemes increase with the transmission power.
However, the performances of SDMA-based methods gradually saturate with increasing power.
Because the MRT does not eliminate interference. Briefly, the powers of both the useful and the interference signals increase with the power, hence the SINR gradually approaches a constant value.
By contrast, the NOMA-based baseline does not suffer from this phenomenon, because the utilization of SIC can eliminate the interference, and the SNR can increase with transmit power.
Finally, the proposed sub-optimal algorithm can approach the performance of the optimal algorithm and outperform other baselines, which validates the effectiveness of the proposed algorithms.

 \begin{figure*}[t] \centering
 \begin{minipage}{.48 \textwidth}
   \centering
   \includegraphics[width=\textwidth ]{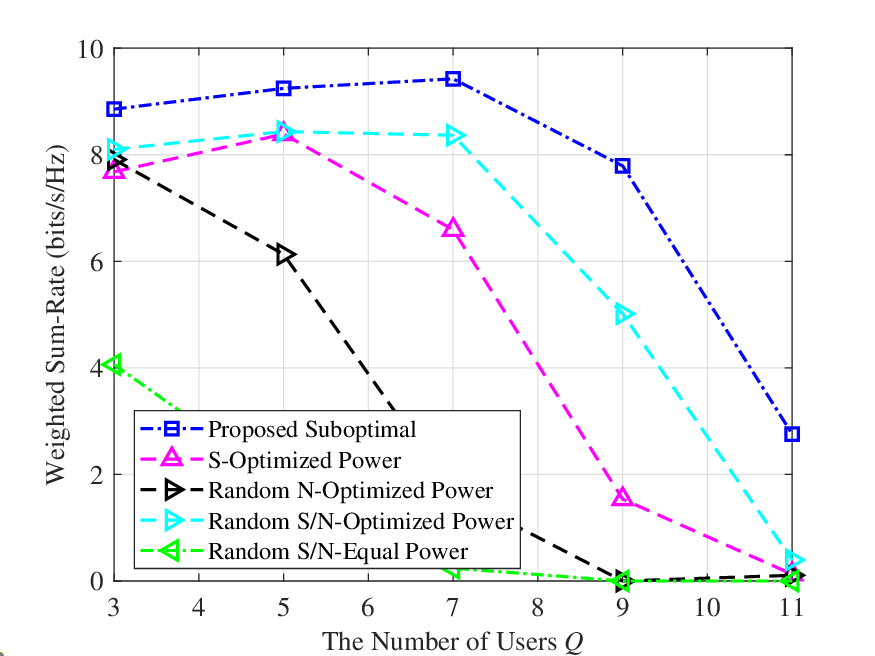}
   \subcaption{Weighted Sum-Rate}\label{figsim31}
  \end{minipage}
   \begin{minipage}{.48 \textwidth}
   \centering
   \includegraphics[width=\textwidth ]{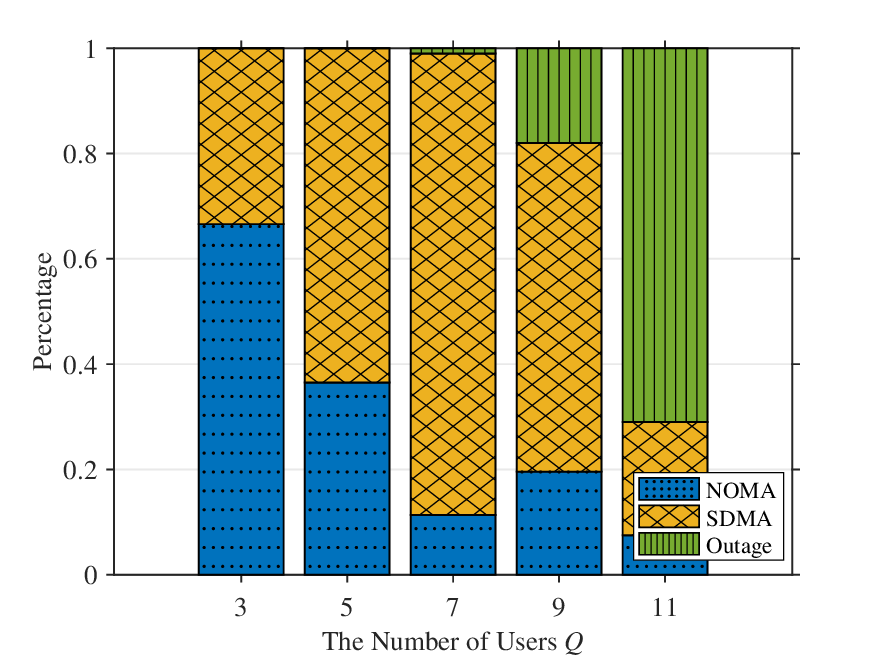}
   \subcaption{Occurrence Percentage}\label{figsim32}
  \end{minipage} 
  \caption{The weighted sum-rate of all users versus the number of users with QoS constraint:  $P=45$ dBm, $C_{\rm min}= 0.6$, $D=8$, $M=3$, $N=8$, $\delta_M=1$,  and $\delta_N=2$.}\label{figsim3}  
  \end{figure*}

  \begin{figure*}[t] \centering
 \begin{minipage}{.48 \textwidth}
   \centering
   \includegraphics[width=\textwidth]{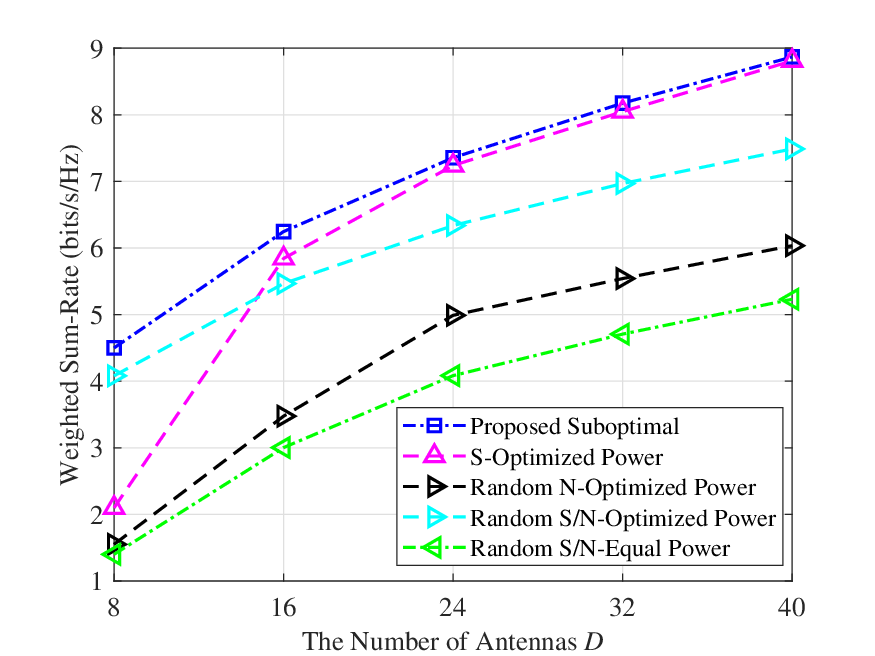}
    \subcaption{Weighted Sum-Rate}\label{figsim41}
  \end{minipage}
   \begin{minipage}{.48  \textwidth}
   \centering
   \includegraphics[width=\textwidth]{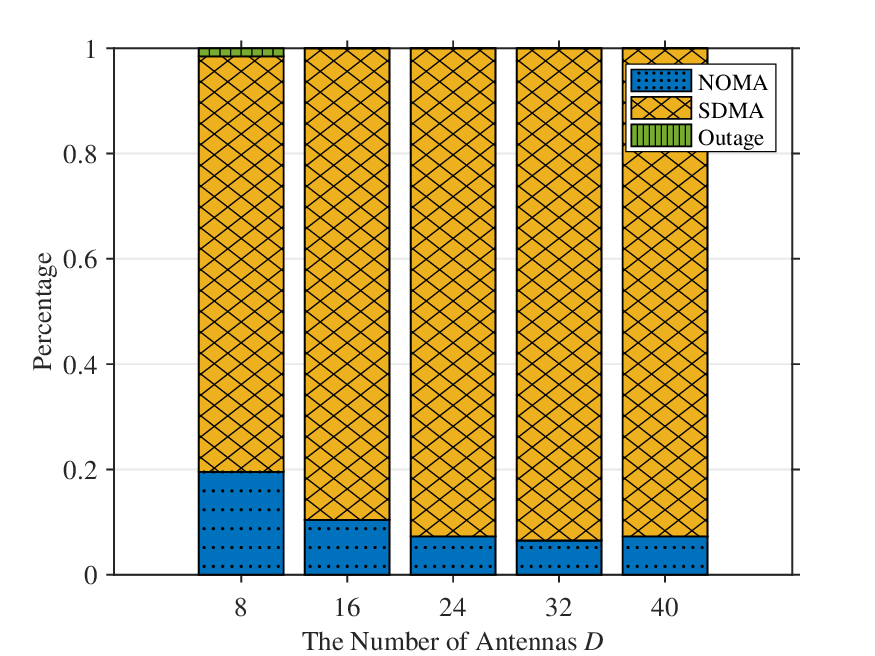}
  \subcaption{Occurrence Percentage}\label{figsim42}
  \end{minipage}
  \caption{The weighted sum-rate of all users versus the number of transmission antennas with QoS constraint:  $P=45$ dBm, $C_{\rm min}= 0.6$, $Q=3$, $M=18$, $N=16$, $\delta_M=6$, and  $\delta_N=8$.}\label{figsim4} 
 \end{figure*}


Fig. \ref{figsim2} investigates the impact of transmission power on the weighted sum-rate of users, while considering the individual rate constraints.
Firstly, we observe that the performances of all schemes degrade due to the rate constraints, especially when the transmission power is low.
Then, the performance of the SDMA-based method is superior to that of the NOMA-based method at low transmission power because the latter struggles to effectively harness SIC. However, at a high transmission power, the performance of SDMA-based methods is limited by the interference, while the NOMA-based method benefits from the increase of power due to harnessing SIC.
Hence, as the power increases, the proposed algorithm gradually shifts from favoring SDMA as the access scheme to favoring NOMA.
Finally, the outage probability is reduced upon increasing the transmission power and the proposed algorithm outperforms the baselines.

Fig. \ref{figsim3} investigates the weighted sum-rate of users versus the number of users.
Firstly, it shows that the performances of most schemes initially increase and then decrease as the number of users increases.
The reason behind this trend is that the increased number of users can provide a higher degree of freedom for resource allocation, thus achieving an improved performance.
However, the resources are limited and each newly added user has an individual minimal transmission rate.
 Hence,  the outage probability will increase due to the increased number of additional users, which further decreases the performance.
Furthermore, the performance of the SDMA-based method is relatively poorer compared to the NOMA-based method, when the number of users is low. However, as the number of users increases, the SDMA-based method exhibits superior performance compared to the NOMA-based method.
This is attributed to the fact that for a smaller user count, the latter can apply SIC to cancel the interference, hence improving the transmission rate, while the latter readily meets the rate requirements without performing SIC for a sufficiently higher number of users.
Therefore,  as the number of users increases, the proposed algorithm gradually shifts from favoring NOMA to favoring SDMA.
 This validates the effectiveness of the proposed algorithms.

Fig. \ref{figsim4} investigates the weighted sum-rate of users versus the number of antennas.
Firstly, it shows that the performances of all schemes increase with the number of antennas.
The reason behind this trend is that the increased number of antennas can achieve higher beamforming gain, hence improving the SINR.
Next, the performance of the SDMA-based method is relatively poorer compared to the NOMA-based method, when the number of antennas is low.
However, as the number of antennas increases, the SDMA-based method exhibits superior performance compared to the NOMA-based method.
Because upon increasing the number of antennas, the channels gradually become orthogonal. Thus, the SDMA-based method can simultaneously provide orthogonal transmissions to multiple users, whereas the NOMA-based method can only serve a maximum of two users. This leads to an increasing performance gap between them.
Therefore, as the number of antennas increases, the proposed algorithm gradually shifts from favoring NOMA as the access scheme to favoring SDMA, which underlines the effectiveness of the advocated algorithm.

\begin{figure}[t]\centering
   \centering
   \includegraphics[width=0.48\textwidth]{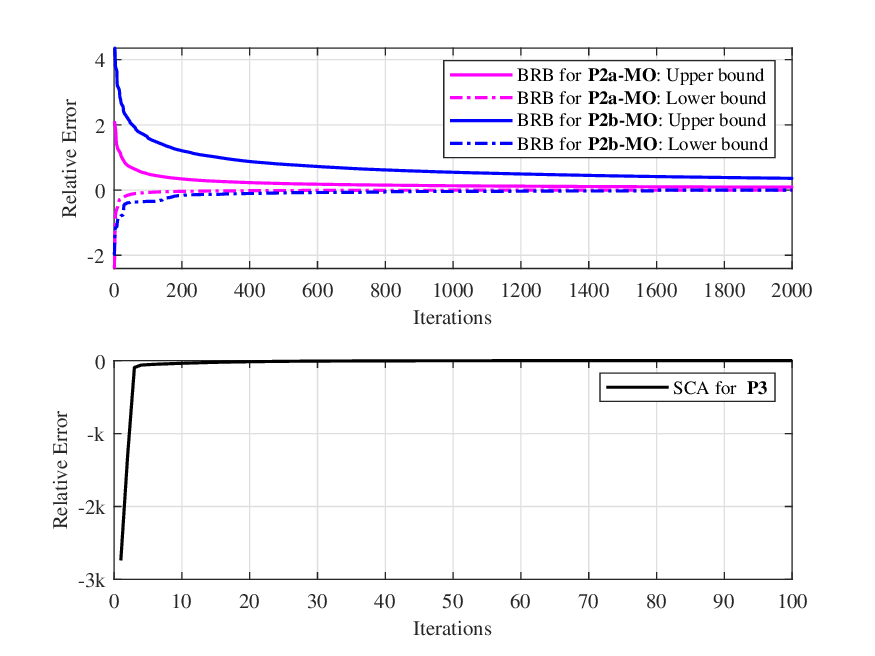}
   \caption{Convergence performances of BRB and SCA algorithms: $P=45$ dBm, $Q=3$, $D=10$, $M=2$, $N=4$, $\delta_M=1$, and $\delta_N=2$.}\label{figconvergence}
 \end{figure}

 Fig. \ref{figconvergence} investigates the convergence performance of the BRB and SCA algorithms.
 Firstly,  the upper and lower bounds of the BRB algorithm gradually approach each other and converge finally. Besides, the convergence speed of BRB in addressing SDMA-based problems is lower than that for NOMA-based problems.
 Because the former involves optimizing a higher number of variables.
 Finally, the SCA converges in just a few iterations, which confirms its low complexity.

\section{Conclusions}
We proposed an elastic multi-domain resource utilization mechanism for mobile communication and introduced a multi-user OTFS-MDMA system, which can leverage the associated user-specific channel characteristics across multi-domain resources including the DD, power, and spatial domains.
Then, we formulated a weighted sum-rate maximization problem by jointly optimizing the user accommodation, access scheme selection, and power allocation subject to individual rate constraints and various other practical constraints.
Since the problem formulated is a non-convex problem, we developed an efficient algorithm based on DPMO to find its globally optimal solution in the special case of disregarding rate constraints.
Subsequently, we design a low-complexity SCA-SA algorithm to find its sub-optimal solutions in the general case.
Finally, the simulation results demonstrate the benefits of employing elastic multi-domain resource utilization.

Note that the proposed sub-optimal algorithm still introduces some complexity that may affect the real-time application of OTFS-MDMA design. A potential solution is to more effectively approximate the original problem formulation and develop a lower-complexity sub-optimal algorithm.
Besides, the proposed OTFS-MDMA design requires channel knowledge of all users.
To reduce estimation overhead, we can incorporate channel temporal correlation and use learning-based methods \cite{chen2024learning} to allocate resources based on historical channel knowledge rather than current perfect knowledge. However, as this paper focuses on the principle and validation of the OTFS-MDMA design, we leave addressing the challenges of complexity and channel estimation overhead reduction in real-time applications as an interesting area for future work.

\appendix

\subsection{Derivation of \eqref{eqdiag}}\label{theorem0proof}

In this part, we first introduce the following lemma in \cite{raviteja2018practical} and then apply it to derive \eqref{eqdiag}.
\begin{lemma} \cite{raviteja2018practical}: Let  ${\bf{G}} = {\rm{circ}}\left[ {{{\bf{G}}_0},{{\bf{G}}_1}, \cdots {{\bf{G}}_{N - 1}}} \right]\in{\mathbb C}^{NM\times NM}$ be a block circulant matrix, where ${\bf G}_n\in{\mathbb C}^{M\times M}$ for $0\le n\le N-1$. Then, ${\bf{G}}$ can be re-expressed by a block diagonal matrix, i.e.,
 \begin{align}
{\bf{G}} &=\left( {{\bf{F}}_N^H \otimes {{\bf{I}}_M}} \right){\rm{diag}}\left( {{{\bm\Theta}_i},0 \le i \le N - 1} \right)\left( {{{\bf{F}}_N} \otimes {{\bf{I}}_M}} \right)\nonumber\\
&=\left( {{{\bf{F}}_N} \otimes {{\bf{I}}_M}} \right) {\rm{diag}}\left( {{{\tilde{\bm \Theta}} _i},0 \le i \le N - 1} \right) \left( {{\bf{F}}_N^H \otimes {{\bf{I}}_M}} \right),\nonumber
\end{align}
where ${{{\bm \Theta} _i} = \sum\limits_{n = 0}^{N - 1} {{{\bf{G}}_n}{e^{ - {\rm{j2}}\pi \frac{{ni}}{{N}}}}} }$ and ${{\tilde {\bm \Theta} _i} = \sum\limits_{n = 0}^{N - 1} {{{\bf{G}}_n}{e^{ {\rm{j2}}\pi \frac{{ni}}{{N}}}}} }$.
\hfill $\square$
\end{lemma}

Based on this lemma,  ${{\bf{H}}_q^d}$ can be represented by
 \begin{align}
&{\bf{H}}_q^d \nonumber\\
= &\left( {{\bf{F}}_N^H \otimes {{\bf{I}}_M}} \right){\rm{diag}}\left( {{\bm \Theta} _{q,0}^d, \cdots ,{\bm \Theta} _{q,N - 1}^d} \right)\left( {{{\bf{F}}_N} \otimes {{\bf{I}}_M}} \right),
\end{align}
where ${{\bm \Theta} _{q,i}^d = \sum\nolimits_{n = 0}^{N - 1} {{\bf{G}}_{q,n}^d{e^{ - {\rm{j2}}\pi \frac{{ni}}{{N}}}},0 \le i \le N - 1} }$. Due to the definition of ${\bf{G}}_{q,n}^d $ in \eqref{eqBqn}, we know ${\bf{G}}_{q,n}^d$ is a sum of $N$ circulant matrices, which means a similar diagonalization processes are present, i.e., ${\bf{G}}_{q,n}^d = {{\bf{F}}_M}{\bm \Lambda}_{q,n}^d{\bf{F}}_M^H$ with ${\bm \Lambda}_{q,n}^d$ defined after \eqref{eqdiag}. Then, we have
\begin{align}
{\bf{H}}_q^d = \left( {{\bf{F}}_N^H \otimes {{\bf{F}}_M}} \right){\bm{\Lambda }}_q^d\left( {{{\bf{F}}_N} \otimes {\bf{F}}_M^H} \right),
 \end{align}
where ${\bm{\Lambda }}_q^d$ is defined after \eqref{eqdiag}.
Finally, we have \eqref{eqdiag}.


\ifCLASSOPTIONcaptionsoff

\fi
  \bibliography{SPT}
\bibliographystyle{IEEEtran}

\begin{IEEEbiography}[{\includegraphics[width=1in,height=1.25in,clip,keepaspectratio]{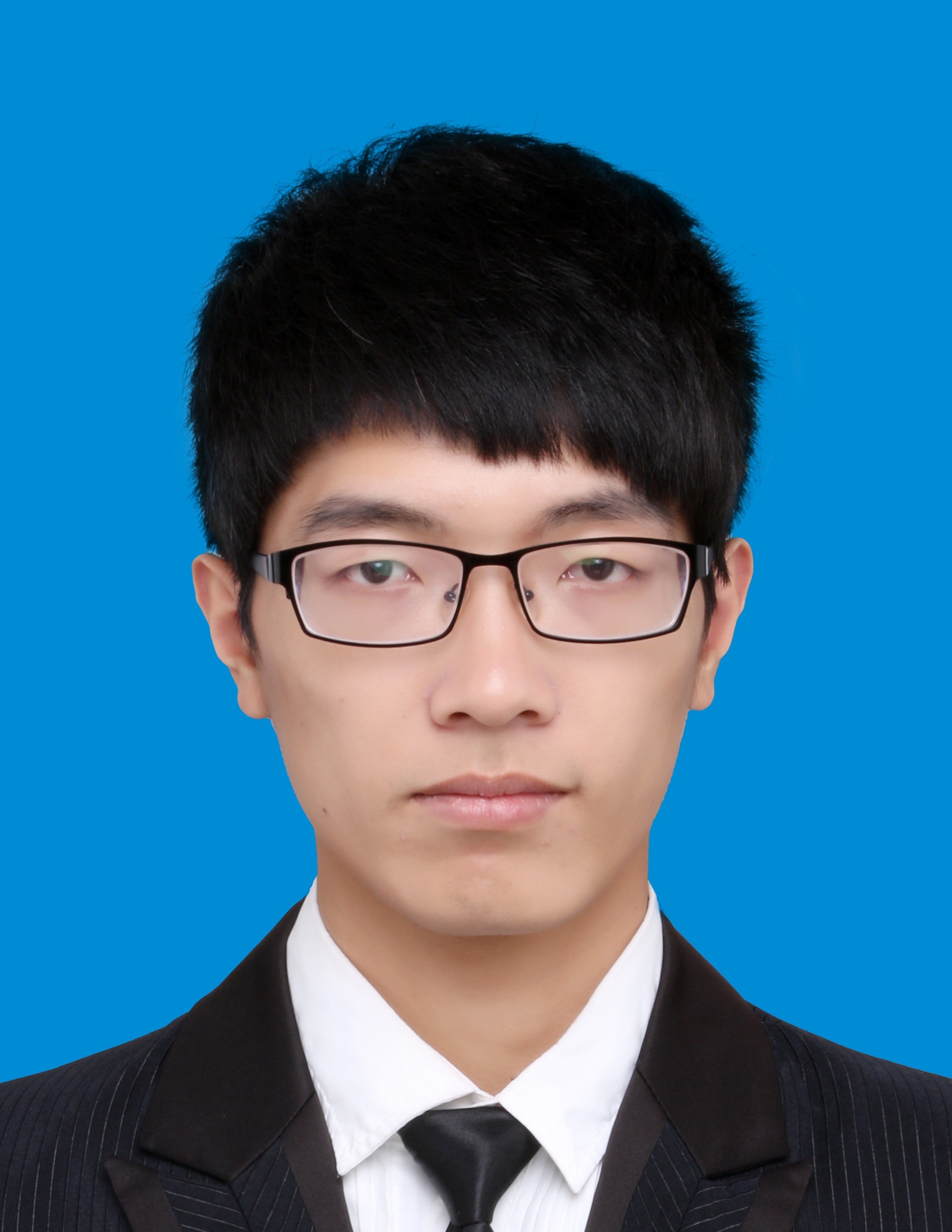}}]{Jie Chen} (Member, IEEE) received the B.S. degree in communication engineering from Chongqing University of Posts and Telecommunications, China, in 2016, and the Ph.D. degree from University of Electronic Science and Technology of China (UESTC), China, in 2021.
From 2019-2020, he was a Visiting Student Research Collaborator at the University of Toronto, Toronto, ON, Canada.
He is currently a Postdoctoral Research Fellow with the Department of Electrical and Computer Engineering, Western University, London, ON, Canada.
His research interests include integrated sensing and communications, transceiver design for Internet-of-Things, and machine learning for wireless communications.
He was the recipient of the Journal of Communications and Information Networks (JCIN) Best Paper Award in 2021.
\end{IEEEbiography}

\begin{IEEEbiography}[{\includegraphics[width=1in,height=1.25in,clip,keepaspectratio]{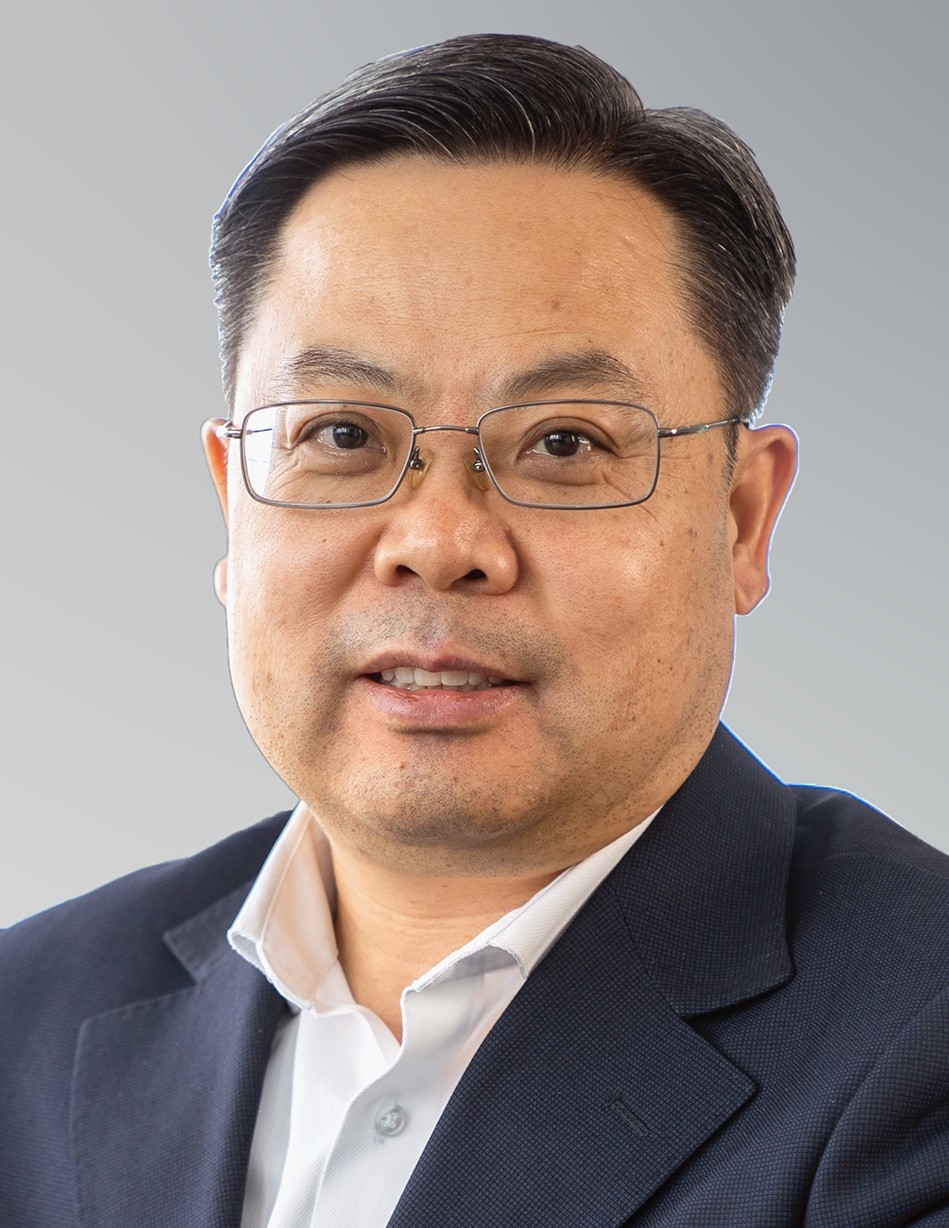}}]{Xianbin Wang} (Fellow, IEEE)  received his Ph.D. degree in electrical and computer engineering from the National University of Singapore in 2001.

He is a Professor and a Tier-1 Canada Research Chair in 5G and Wireless IoT Communications with Western University, Canada. Prior to joining Western University, he was with the
Communications Research Centre Canada as a Research Scientist/Senior Research Scientist from 2002 to 2007. From 2001 to 2002, he was a System Designer at STMicroelectronics. His current research interests include 5G/6G technologies, Internet of Things, machine learning, communications security, and intelligent communications. He has over 600 highly cited journals and conference papers, in addition to over 30 granted and pending patents and several standard contributions.

Dr. Wang is a Fellow of the Canadian Academy of Engineering and a Fellow of the Engineering Institute of Canada. He has received many prestigious awards and recognitions, including the IEEE Canada R. A. Fessenden Award, Canada Research Chair, Engineering Research Excellence Award at Western University, Canadian Federal Government Public Service Award, Ontario Early Researcher Award, and nine Best Paper Awards. He is currently a member of the Senate, Senate Committee on Academic Policy and Senate Committee on University Planning at Western. He also serves on NSERC Discovery Grant Review Panel for Computer Science. He has been involved in many flagship conferences, including GLOBECOM, ICC, VTC, PIMRC, WCNC, CCECE, and ICNC, in different roles, such as General Chair, TPC Chair, Symposium Chair, Tutorial Instructor, Track Chair, Session Chair, and Keynote Speaker. He serves/has served as the Editor-in-Chief, Associate Editor-in-Chief, and editor/associate editor for over ten journals. He was the Chair of the IEEE ComSoc Signal Processing and Computing for Communications (SPCC) Technical Committee and is currently serving as the Central Area
Chair of IEEE Canada.

\end{IEEEbiography}

\begin{IEEEbiography}[{\includegraphics[width=1in,height=1.25in,clip,keepaspectratio]{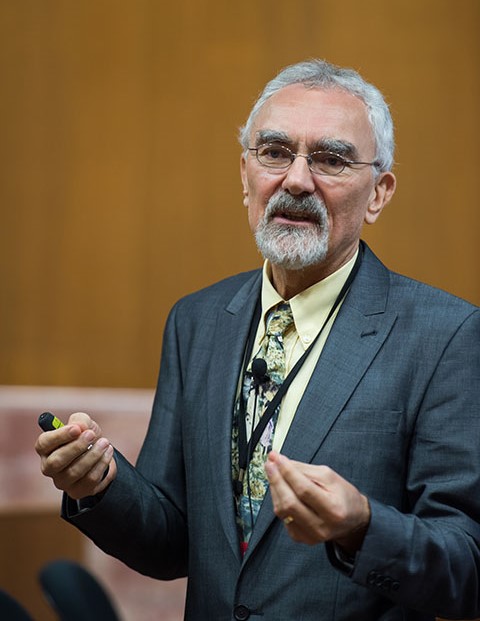}}]{Lajos Hanzo} (Life Fellow, IEEE) received Honorary Doctorates  from the Technical University of Budapest (2009) and Edinburgh University (2015). He is a Foreign Member of the Hungarian Science-Academy, Fellow of the Royal Academy of Engineering (FREng), of the IET, of EURASIP and holds the IEEE Eric Sumner Technical Field Award.  For further details please see \url{http://www-mobile.ecs.soton.ac.uk, https://en.wikipedia.org/wiki/Lajos_Hanzo}

He coauthored 2000+ contributions at IEEE Xplore and 19 Wiley-IEEE Press monographs. Furthermore, he advised 125 PhD students and about 50 of them are now professors at various stages of their careers.

\end{IEEEbiography}

\end{document}